\documentclass[prd,twocolumn,showpacs,superscriptaddress,nofootinbib,floatfix,showkeys,10pt]{revtex4-2}
\usepackage{bm}
\usepackage{times}
\usepackage{braket}
\usepackage{amsfonts,amssymb,stmaryrd,latexsym,amsmath}
\usepackage[usenames,dvipsnames]{color}
\usepackage{epsfig}
\usepackage{slashed}
\usepackage{hyperref}
\usepackage{subfigure}
\usepackage{graphicx}
\usepackage{slashed}
\usepackage{orcidlink}
\usepackage{multirow}
\usepackage{appendix}
\usepackage{dashrule}

\allowdisplaybreaks[1]

\definecolor{nicered}{rgb}{0.7,0.1,0.1}
\definecolor{nicegreen}{rgb}{0.1,0.5,0.1}
\definecolor{emph}{rgb}{1,0,0}
\definecolor{doub}{rgb}{0.7,0.2,1.0}
\definecolor{navyblue}{RGB}{0, 110, 184}

\hypersetup{colorlinks,citecolor=nicegreen,linkcolor=nicered,urlcolor=navyblue}

 \newcommand{\clabel}[2][]{#2}
 \newcommand{\change}[1]{#1}
\begin{document}
%
	
	\title{Benchmark calculations of fully heavy compact and molecular tetraquark states} 
	\author{Wei-Lin Wu\,\orcidlink{0009-0009-3480-8810}}\email{wlwu@pku.edu.cn}
	\affiliation{School of Physics, Peking University, Beijing 100871, China}
	\author{Yan-Ke Chen\,\orcidlink{0000-0002-9984-163X}}\email{chenyanke@stu.pku.edu.cn}
	\affiliation{School of Physics, Peking University, Beijing 100871, China}
	\author{Lu Meng\,\orcidlink{0000-0001-9791-7138}}\email{lu.meng@rub.de}
	\affiliation{Institut f\"ur Theoretische Physik II, Ruhr-Universit\"at Bochum,  D-44780 Bochum, Germany }
	\author{Shi-Lin Zhu\,\orcidlink{0000-0002-4055-6906}}\email{zhusl@pku.edu.cn}
	\affiliation{School of Physics and Center of High Energy Physics,
		Peking University, Beijing 100871, China}
	
	\begin{abstract}
		We calculate the mass spectrum of the S-wave fully heavy tetraquark systems $ QQ\bar Q\bar Q~(Q=c,b) $ with both normal $ (J^{PC}=0^{++},1^{+-},2^{++}) $ and exotic $ (J^{PC}=0^{+-},1^{++},2^{+-}) 
$ C-parities using three different quark potential models (AL1, AP1, BGS). The exotic C-parity systems refer to the ones that cannot be composed of two S-wave ground heavy quarkonia. We incorporate the molecular dimeson and compact diquark-antidiquark spatial correlations simultaneously, thereby discerning the actual configurations of the states. We employ the Gaussian expansion method to solve the four-body Schrödinger equation, and the complex scaling method to identify the resonant states. The mass spectra in three different models qualitatively agree with each other. We obtain several resonant states with $ J^{PC} = 0^{++}, 1^{+-}, 2^{++}, 1^{++} $ in the mass region $(6.92,7.30)\, \mathrm{GeV}$, some of which are good candidates of the experimentally observed $X(6900)$ and $X(7200)$. We also obtain several exotic C-parity zero-width states with $ J^{PC}=0^{+-} $ and $ 2^{+-} $. These zero-width states have no corresponding S-wave diquarkonium threshold and can only decay strongly to final states with P-wave quarkonia. With the notation $T_{4Q,J(C)}(M)$, we deduce from the root mean square radii that the $ X(7200) $ candidates $ T_{4c,0(+)}(7173), T_{4c,2(+)}(7214) $ and the state $ T_{4c,1(-)}(7191) $ look like molecular states although most of the resonant and zero-width states are compact states.

	\end{abstract}
	
	\maketitle
	
	\section{Introduction}~\label{sec:intro}
	Since the discovery of $X(3872)$~\cite{Belle:2003nnu}, numerous candidates for multiquark states have been observed in experiments. The multiquark states exhibit more intricate color structures in forming color-singlet states than the conventional hadrons. Investigating these states can enhance our understanding of quantum chromodynamics (QCD). One can find more details in recent reviews~\cite{Chen:2016qju,Esposito:2016noz,Hosaka:2016pey,Lebed:2016hpi,Guo:2017jvc,Ali:2017jda,Brambilla:2019esw,Liu:2019zoy,Meng:2022ozq,Mai:2022eur,Chen:2022asf}.

	Among the myriad multiquark systems, the fully heavy tetraquark states $QQ\bar{Q}\bar{Q}~(Q=b, c)$ stand out as relatively pure and clear systems, unaffected by unquenched dynamics such as the creation and annihilation of light $q\bar{q}~(q=u,d,s)$ pairs. Recently, significant progress has been made in the search for fully heavy tetraquark states. The LHCb Collaboration initially discovered a fully charmed tetraquark candidate $X(6900)$~\cite{LHCb:2020bwg}. Subsequently, both the CMS~\cite{CMS:2023owd} and ATLAS~\cite{ATLAS:2023bft} collaborations independently verified the existence of the $X(6900)$ state and reported additional fully charmed tetraquark resonant states. Specifically, the CMS reported the observation of $X(6600)$ and the evidence of $X(7200)$~\cite{CMS:2023owd}, while the ATLAS reported the evidence of $X(6400)
$, $X(6600)$, and $X(7200)$~\cite{ATLAS:2023bft}. Theoretical studies on fully heavy tetraquark states began long before the experiments, making predictions of their existence~\cite{Iwasaki:1975pv,Chao:1980dv,Ader:1981db,Zouzou:1986qh,Heller:1986bt,Silvestre-Brac:1992kaa,Silvestre-Brac:1993wyf,Wu:2016vtq,Chen:2016jxd,Wang:2017jtz,Richard:2017vry,Wang:2019rdo,Bedolla:2019zwg}. 
	After recent discoveries, many theoretical efforts have been made to understand the experimental results~\cite{Deng:2020iqw,liu:2020eha,Jin:2020jfc,Lu:2020cns,Chen:2020xwe,Wang:2020gmd,Albuquerque:2020hio,Giron:2020wpx,Wang:2020wrp,Dong:2020nwy,Zhang:2020hoh,Zhao:2020nwy,Gordillo:2020sgc,Weng:2020jao,Zhang:2020xtb,Guo:2020pvt,Gong:2020bmg,Wan:2020fsk,Dosch:2020hqm,Yang:2020wkh,Huang:2020dci,Zhao:2020jvl,
	Hughes:2021xei,Faustov:2021hjs,Ke:2021iyh,Liang:2021fzr,Yang:2021hrb,Mutuk:2021hmi,Li:2021ygk,Wang:2021kfv,Dong:2021lkh,Wang:2021mma,Liu:2021rtn,Zhuang:2021pci,Asadi:2021ids,
	Kuang:2022vdy,Gong:2022hgd,Wang:2022jmb,Zhou:2022xpd,Chen:2022sbf,An:2022qpt,Wang:2022yes,Niu:2022cug,Zhang:2022qtp,Dong:2022sef,Yu:2022lak,
	Ortega:2023pmr,Wang:2023jqs,Sang:2023ncm,Galkin:2023wox,Anwar:2023fbp}. The interpretations of these states include compact tetraquark states~\cite{Deng:2020iqw,liu:2020eha,Jin:2020jfc,Lu:2020cns,Giron:2020wpx,Gordillo:2020sgc,Yang:2020wkh,Huang:2020dci,Zhao:2020jvl,Faustov:2021hjs,Ke:2021iyh,Mutuk:2021hmi,Li:2021ygk,Wang:2021kfv,Wang:2021mma,Liu:2021rtn,Asadi:2021ids,An:2022qpt,Zhang:2022qtp,Dong:2022sef,Yu:2022lak,Galkin:2023wox}, dynamical effects in dicharmonium rescattering~\cite{Wang:2020wrp,Dong:2020nwy,Guo:2020pvt,Gong:2020bmg,Liang:2021fzr,Zhuang:2021pci,Gong:2022hgd,Wang:2022jmb,Zhou:2022xpd,Ortega:2023pmr}, hybrid states~\cite{Wan:2020fsk}, etc. Further details can be found in recent reviews~\cite{Mai:2022eur,Chen:2022asf}.
	
	The nature of fully charmed tetraquark states remains controversial. Currently, few theoretical works consider compact diquark-antidiquark and molecular diquarkonium spatial configurations simultaneously and perform comprehensive four-body dynamical calculations. In our previous studies~\cite{Ma:2022vqf,Ma:2023int,Meng:2023jqk}, we incorporated both dimeson and diquark-antidiquark spatial configurations, employing various quark models and few-body methods to conduct benchmark calculations for tetraquark bound states. We have illustrated that the Gaussian expansion method (GEM) is highly efficient in exploring tetraquark states. Our results indicate that there are no bound states in the fully heavy tetraquark systems. In Refs.~\cite{Wang:2022yes,Wang:2023jqs}, the authors utilized the GEM to conduct four-body dynamic calculations for the fully charmed tetraquark systems. Furthermore, they employed the complex scaling method (CSM) \cite{Aguilar:1971ve,Balslev:1971vb,aoyama2006complex} to distinguish resonant states from dicharmonium scattering states, yielding convincing and intriguing results.  However, it is noteworthy that the discussion in Refs.~\cite{Wang:2022yes,Wang:2023jqs} is limited to fully charmed tetraquark states, and does not include an exploration of fully bottomed tetraquark states. Furthermore, the discussion excludes the systems with exotic C-parity $(J^{PC}=0^{+-},1^{++},2^{+-})$, which refer to the ones that cannot be composed of two S-wave ground  heavy quarkonia.
	
	In this study, we aim to investigate the S-wave fully heavy tetraquark $QQ\bar Q\bar Q~(Q=b,c)$ resonances with all possible quantum numbers, employing a framework that has been used to investigate $Qs \bar q \bar q$ states efficiently~\cite{Chen:2023syh}. We employ the GEM to solve the four-body Schrödinger equation, considering both compact diquark-antidiquark and molecular diquarkonium spatial configurations. We utilize the CSM to distinguish resonant states from diquarkonium scattering states. Regarding the (anti)quark-(anti)quark interactions, we compare three quark potential models with well-determined parameters and do not introduce any new free parameters. Additionally, we apply the approach proposed in our previous work~\cite{Chen:2023syh} to analyze the spatial structures of the tetraquark states, which can clearly distinguish between the compact tetraquark states and the molecular states.
	
	This paper is organized as follows. In Sec.~\ref{sec:theo_framwork}, we provide an introduction to the theoretical framework, including (anti)quark-(anti)quark interactions, calculation methods, and the approach to discerning between molecular and compact tetraquark configurations. In Sec.~\ref{sec:result}, we comprehensively discuss our numerical results, exploring the masses, widths, decays, and spatial structures of the fully heavy tetraquark states. We also explore the properties of the states with exotic C-parity. Finally, we summarize our findings in Sec.~\ref{sec:summary}.
	
	\section{Theoretical Framework}\label{sec:theo_framwork}
	
	\subsection{Hamiltonian}\label{subsec:hamiltonian}
	
	In the center-of-mass frame, the nonrelativistic Hamiltonian for a tetraquark system reads
	\begin{equation}
		H=\sum_{i=1}^4 (m_i+\frac{p_i^2}{2 m_i})+\sum_{i<j=1}^4 V_{ij
		},
	\end{equation}
	where $m_i$ and $p_i$ are the mass and momentum of the (anti)quark $i$, respectively. $V_{ij}$ represents the two-body interaction between the (anti)quark pair $(ij)$. In this study, we adopt three different quark potential models, namely the AL1 and AP1 potentials proposed in Refs.~\cite{Semay:1994ht, Silvestre-Brac:1996myf} and the potential used to study charmonia in Ref.~\cite{Barnes:2005pb} (denoted as the BGS potential hereafter). These three potentials contain the one-gluon-exchange interaction and quark confinement interaction, and can be written as
	\begin{equation}
		\label{eq:potential}
		\begin{aligned}
			V_{i j} =-\frac{3}{16} \boldsymbol\lambda_i \cdot \boldsymbol\lambda_j\left(-\frac{\kappa}{r_{i j}}+\lambda r_{i j}^p-\Lambda\right. \\
			\left.+\frac{8 \pi \kappa^{\prime}}{3 m_i m_j} \frac{\exp \left(-r_{i j}^2 / r_0^2\right)}{\pi^{3 / 2} r_0^3} \boldsymbol{S}_i \cdot \boldsymbol{S}_j\right),
		\end{aligned}
	\end{equation}
	where $ r_{ij} $ is the distance between (anti)quark $ i $ and $ j $, $\boldsymbol\lambda_i$ and $ \boldsymbol{S}_i $ are the $\mathrm{SU}(3)$ color Gell-Mann matrix and the spin operator acting on (anti)quark $ i $. The first term in the potential is referred to as the color electric term, and the last term is known as the color magnetic term. In the conventional meson and baryon systems, the color factor $\boldsymbol\lambda_i \cdot \boldsymbol\lambda_j$ is always negative and induces a confining interaction, thus all eigenstates of the Hamiltonian must be bound states. However, scattering states of two color-singlet clusters and possible resonant states are allowed in the tetraquark systems, since they have richer inner color structures than the conventional hadrons, and the color factor might take a zero or positive value. The parameters of the models are taken from Refs.~\cite{Silvestre-Brac:1996myf, Barnes:2005pb} and listed in Table~\ref{tab:para}. The parameters for the AL1 and AP1 potential were determined by fitting the meson spectra across all flavor sectors, while those for the BGS potential were determined by the charmonium spectra. The theoretical masses of the charmonia and bottomonia as well as their root-mean-square (rms) radii calculated from the AP1 potential are listed in Table~\ref{tab:meson}. It can be seen that all three potential models can give a satisfactory description of the meson spectra.
	\begin{table}[htbp]
		\centering
		\caption{The parameters in the AL1, AP1 and BGS quark potential models.}
		\label{tab:para}
		\begin{tabular*}{\hsize}{@{}@{\extracolsep{\fill}}cccc@{}}
			\hline\hline
			Parameters &AL1~\cite{Silvestre-Brac:1996myf}&AP1~\cite{Silvestre-Brac:1996myf}&BGS~\cite{Barnes:2005pb}\\
			\hline
			$ p $&1&$ \frac{2}{3} $&1\\
			$ \kappa $&0.5069&0.4242&0.7281\\
			$ \lambda { [\mathrm{GeV}^{p+1}]} $&0.1653&0.3898&0.1425\\
			$ \Lambda {\rm [GeV]} $&0.8321&1.1313&0\\
			$ \kappa^\prime $&1.8609&1.8025&0.7281\\
			$ m_c {\rm [GeV]}$&1.8360&1.8190&1.4794\\
			$ m_b {\rm [GeV]}$&5.227&5.206&-\\
			$ r_{0c} {\rm [GeV^{-1}]}$&1.4478&1.2583&0.9136\\
			$ r_{0b} {\rm [GeV^{-1}]}$&1.1497&0.8928&-\\
			
			\hline\hline
		\end{tabular*}
	\end{table}
	
	\begin{table}
		\centering
		\caption{The masses (in  $\mathrm{MeV}$) of $ c\bar{c} $ and $ b\bar b $ quarkonia in three different quark models, compared with the experimental results taken from Ref.~\cite{ParticleDataGroup:2022pth}. The rms radii (in fm) of the quarkonia in the AP1 potential are listed in the last column. }
		\label{tab:meson}
		\begin{tabular*}{\hsize}{@{}@{\extracolsep{\fill}}cccccc@{}}
			\hline\hline
			Mesons& $ m_{\rm Exp.} $&$ m_{\rm AL1} $ &$ m_{\rm AP1} $ &$ m_{\rm BGS} $&$ r^{\rm rms}_{\rm AP1} $ \\
			\hline
			$\eta_c $ &2984& 3006 & 2982&2982&0.35\\
			$\eta_c(2S) $ &3638& 3608 & 3605&3630&0.78\\
			$\eta_c(3S) $ &-& 4014 & 3986&4043&1.15\\
			$J/\psi $ &3097& 3102 & 3102&3090&0.40\\
			$\psi(2S) $ &3686& 3641 & 3645&3672&0.81\\
			$\psi(3S) $&4039& 4036 & 4011&4072&1.17\\
			$\eta_b $ &9399& 9424 & 9401 &-&0.20\\
			$\eta_b(2S) $&9999 & 10003 & 10000&-&0.48\\
			$\eta_b(3S) $&-& 10329 & 10326 &-&0.73\\
			$\Upsilon $ &9460& 9462 & 9461 &-&0.21\\
			$\Upsilon(2S) $&10023 & 10012 & 10014 & -&0.49\\
			$\Upsilon(3S) $&10355& 10335 & 10335 & -&0.74\\
			
			\hline\hline
		\end{tabular*}
	\end{table}
	
	\subsection{Calculation methods}\label{subsec:the_calculation_method}
	We use the complex scaling method (CSM) to obtain possible bound and resonant states simultaneously, and we apply the Gaussian expansion method (GEM) to solve the complex-scaled four-body Schrödinger equation. We also introduce a method to determine the C-parity of the neutral tetraquark states by decomposing the Hilbert space.
	
	In the CSM~\cite{Aguilar:1971ve, Balslev:1971vb,aoyama2006complex}, the coordinate $ \boldsymbol{r} $ and its conjugate momentum $ \boldsymbol{p} $ are  transformed as
	\begin{equation}
		U(\theta) \boldsymbol{r}=\boldsymbol{r} e^{i \theta}, \quad U(\theta) \boldsymbol{p}=\boldsymbol{p} e^{-i \theta}.
	\end{equation}
	Under such a transformation, the complex-scaled Hamiltonian is written as
	\begin{equation}
		H(\theta)=\sum_{i=1}^4 (m_i+\frac{p_i^2e^{-2i\theta}}{2 m_i})+\sum_{i<j=1}^4 V_{ij}(r_{ij}e^{i\theta}),
	\end{equation}
	which is no longer hermitian and has complex eigenvalues $ E(\theta) $. According to the ABC theorem~\cite{Aguilar:1971ve,Balslev:1971vb}, the energies of the bound states, resonant states and scattering states can be obtained as the eigenvalues of $ H(\theta) $ simultaneously. The bound states (zero-width states) lie on the negative real axis in the energy plane and are not changed by the complex scaling. The resonant states with mass $ M_R $ and width $ \Gamma_R $ can be detected at $ E_R=M_R-i\Gamma_R/2 $ when $ 2\theta>\left|\operatorname{Arg}(E_R)\right| $. The scattering states line up along beams starting from threshold energies and rotated clockwise by $ 2\theta $ from the positive real axis.

\begin{figure}[htbp]
		\centering
		\includegraphics[width=.9\linewidth]{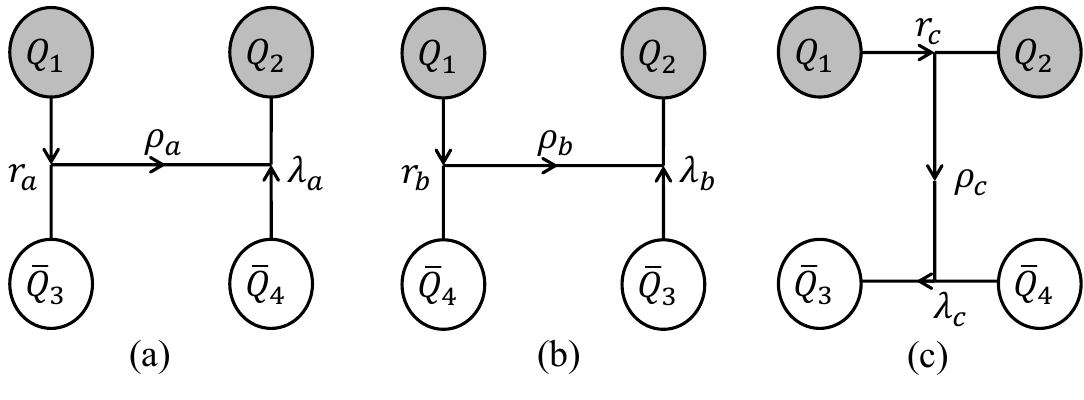}
		\caption{The Jacobian coordinates for two types of spatial configurations: (a), (b) for the dimeson configurations, and (c) for the diquark-antidiquark configuration.}
		\label{fig:jac}
	\end{figure}

	To solve the complex-scaled Schrödinger equation, we apply the GEM~\cite{HIYAMA2003223} and expand the wave functions of the S-wave fully heavy tetraquark states with total angular momentum $ J $ and C-parity $ C $ as
	
	\begin{equation}\label{eq:wavefunction}
		\begin{aligned}
			\Psi^{J}_C(\theta)=\mathcal{A}\sum_{\rm{jac}}\sum_{\alpha,n_{i}}\,C^{(\rm{jac})}_{\alpha,n_{i}}(\theta)\left[\chi_\alpha^J\Phi^{(\rm{jac})}_{n_{1},n_{2},n_{3}}\right]_C.
		\end{aligned}
	\end{equation}
	where $ \mathcal{A} $ is the antisymmetric operator of identical particles. $ \chi_\alpha^J $ is the color-spin wave function, given by
	\begin{equation}\label{eq:colorspin_wf1}
		\begin{aligned}
			\chi^{J}_{\bar 3_c\otimes 3_c,s_1,s_2}=\left[\left(Q_1Q_2\right)_{\bar 3_c}^{s_1}\left(\bar Q_3\bar Q_4\right)_{3_c}^{s_2}\right]_{1_c}^{J},\\
			\chi^{J}_{6_c\otimes \bar 6_c,s_1,s_2}=\left[\left(Q_1Q_2\right)_{6_c}^{s_1}\left(\bar Q_3\bar Q_4\right)_{\bar{6}_c}^{s_2}\right]_{1_c}^{J},\\
		\end{aligned}
	\end{equation}
	for all possible combinations of $ s_1, s_2, J $. It should be emphasized that one has the flexibility to use either the diquark-antidiquark type color-spin basis in Eq.~\eqref{eq:colorspin_wf1} or the dimeson type color-spin basis, which can be written as
	\begin{equation}\label{eq:colorspin_wf2}
		\begin{aligned}
			\chi^J_{1_c\otimes 1_c,s_1,s_2}=\left[\left(Q_1\bar Q_3\right)^{s_1}_{1_c}\left(Q_2\bar Q_4\right)^{s_2}_{1_c}\right]_{1_c},\\
			\chi^J_{8_c\otimes 8_c,s_1,s_2}=\left[\left(Q_1\bar Q_3\right)^{s_1}_{8_c}\left(Q_2\bar Q_4\right)^{s_2}_{8_c}\right]_{1_c}.\\
		\end{aligned}
	\end{equation}
	These two sets of bases are both complete and the transformation between them are shown in Appendix~\ref{app:color}. It has been demonstrated that the use of different discrete basis functions yields negligible differences once they are complete~\cite{Meng:2023jqk}. The S-wave spatial wave function $ \Phi^{(\rm{jac})}_{n_{1},n_{2},n_{3}} $ is written as

	\begin{equation}\label{eq:spatial_wf}
		\Phi^{(\rm{jac})}_{n_{1},n_{2},n_{3}} = \phi_{n_{1}}(r_{jac})\phi_{n_{2}}(\lambda_{jac})\phi_{n_{3}}(\rho_{jac})
	\end{equation}
	where $ (\rm{jac}) = (a),\,(b),\,(c) $  denotes three sets of spatial configurations (dimeson and diquark-antidiquark) considered in our calculations, and $r_{jac}, \lambda_{jac}$, $\rho_{jac}$ are three independent Jacobian coordinates in configuration $ (\rm{jac}) $, as shown in Fig.~\ref{fig:jac}.  $ \phi_{n_i}(r) $ takes the Gaussian form,
	\begin{equation}
		\begin{array}{c}
			\phi_{n_i}(r)=N_{n_{i}}e^{-\nu_{n_i}r^2},\\
			\nu_{n_i}=\nu_{1}\gamma^{n_i-1},
		\end{array}
	\end{equation}
	where $ N_{n_i} $ is the normalization factor. Finally, the expansion coefficients $ C^{(\rm{jac})}_{\alpha,n_i}(\theta) $ are determined by solving the energy eigenvalue equation,
	\begin{equation}
		H(\theta)\Psi_C^J(\theta)=E(\theta)\Psi_C^J(\theta).
	\end{equation}
	
	For the fully heavy neutral tetraquark system $ (QQ\bar Q\bar Q) $, the eigenstates of the Hamiltonian have definite C-parity $ C $. To determine the C-parity of the obtained states, we decompose the Hilbert space $ \mathcal{H} $ into $ \mathcal{H}_+ $ and $ \mathcal{H}_- $, where $ \mathcal{H}_\pm $ represents the subspace with $ C = \pm1 $. Under charge conjugation, the basis functions with $ (\rm{ jac})=(a)$ are transformed as
	\begin{equation}\label{eq:Ctransform1}
		\begin{aligned}
			&[(Q_1Q_2)^{s_1}_{x_c}(\bar{Q}_3\bar{Q}_4)^{s_2}_{\bar{x}_c}]^{J}_{1_c}\Phi^{(\rm{a})}_{n_{1},n_{2},n_{3}}\\
			\stackrel{\mathbf{C}}{\rightarrow}&[(\bar Q_1\bar Q_2)^{s_1}_{\bar x_ c}({Q}_3{Q}_4)^{s_2}_{x_c}]^{J}_{1_c}\Phi^{(\rm{a})}_{n_{1},n_{2},n_{3}}\\
			=&(-1)^{s_1+s_2-J}[({Q}_3{Q}_4)^{s_2}_{{x_c}}(\bar Q_1\bar Q_2)^{s_1}_{\bar  x_c}]^{J}_{1_c}\Phi^{(\rm{a})}_{n_{1},n_{2},n_{3}}\\
			=&(-1)^{s_1+s_2-J}[(Q_1Q_2)^{s_2}_{{x_c}}(\bar Q_3\bar Q_4)^{s_1}_{\bar  x_c}]^{J}_{1_c}\Phi^{(\rm{a})}_{n_{1},n_{2},n_{3}}.
		\end{aligned}
	\end{equation}
	In the third line, we exchange $ [{Q}_3{Q}_4]^{s_2}_{x_c} $ and $ [\bar Q_1\bar Q_2]^{s_1}_{\bar x_c} $ in the color-spin wave function and the factor $ (-1)^{s_1+s_2-J} $ arises from the Clebsch–Gordan coefficient. In the last line the particle indices $ 1\leftrightarrow3, 2\leftrightarrow4 $ are exchanged to rewrite the basis function to its original form. The transformation behavior of the basis functions with $(\rm{jac})=(b),(c)$ can be obtained similarly,
	\begin{equation}\label{eq:Ctransform2}
		\begin{aligned}
			&[(Q_1Q_2)^{s_1}_{x_c}(\bar{Q}_3\bar{Q}_4)^{s_2}_{\bar x_{c}}]^{J}_{1_c}\Phi^{(\rm{b})}_{n_{1},n_{2},n_{3}}\\
			\stackrel{\mathbf{C}}{\rightarrow}
			&(-1)^{s_1+s_2-J}[(Q_1Q_2)^{s_2}_{{x_c}}(\bar Q_3\bar Q_4)^{s_1}_{\bar x_c}]^{J}_{1_c}\Phi^{(\rm{b})}_{n_{2},n_{1},n_{3}},
		\end{aligned}
	\end{equation}
	\begin{equation}\label{eq:Ctransform3}
		\begin{aligned}
			&[(Q_1Q_2)^{s_1}_{x_c}(\bar{Q}_3\bar{Q}_4)^{s_2}_{\bar x_{c}}]^{J}_{1_c}\Phi^{(\rm{c})}_{n_{1},n_{2},n_{3}}\\
			\stackrel{\mathbf{C}}{\rightarrow}
			&(-1)^{s_1+s_2-J}[(Q_1Q_2)^{s_2}_{x_{c}}(\bar Q_3\bar Q_4)^{s_1}_{\bar  x_c}]^{J}_{1_c}\Phi^{(\rm{c})}_{n_{2},n_{1},n_{3}}.
		\end{aligned}
	\end{equation}
	For these two sets of spatial configurations, the indices of Gaussian basis $ (n_{1}, n_{2}) $ are swapped. Once we obtain the transformation properties, we can construct the basis of $ \mathcal{H}_\pm $ by using linear superposition of the original basis functions. The basis functions of $ \mathcal{H}_\pm $ that satisfy the antisymmetrization of identical fermions are listed in Appendix~\ref{app:basis}. The Hamiltonian is block-diagonal in $ \mathcal{H}=\mathcal{H}_+\oplus\mathcal{H}_- $ because of the conservation of C-parity, and we can obtain states with $ C=\pm1 $ by solving the Schrödinger equation in $ \mathcal{H}_\pm $ separately.
	
	\subsection{Discern between molecular and compact tetraquark states}\label{subsec:spatial_distribution}

\clabel[pcolor]{In contrast to other frameworks, such as those discussed in Refs.~\cite{Meng:2019ilv,Meng:2021kmi}, 
the quark model does not require a prior assumptions regarding the structures of a multiquark state. Basically, the molecular or compact tetraquark states can be discerned through the analysis of their wave functions in the quark model. The proportions of color components and root-mean-square (rms) radii are  commonly used criteria~\cite{Yang:2019itm,Wang:2019rdo,Chen:2023syh}, which reflect the color structure and the spatial structure, respectively. }

\change{However, identifying molecular states based on the proportions of color components may be ambiguous and misleading in the systems with identical particles. An ideal loose molecular state consists of two colorless subclusters, which are widely separated. Due to the antisymmetrization of identical particles, its wave function  takes the form
\begin{equation}
\label{eq:wf_mt}
\begin{aligned}
|\Psi\rangle&=[(Q_1\bar Q_3)_{1_c}(Q_2\bar Q_4)_{1_c}]_{1_c}\otimes |\psi_1\rangle\\
&+[(Q_1\bar Q_4)_{1_c}(Q_2\bar Q_3)_{1_c}]_{1_c}\otimes |\psi_2\rangle\\
&=\mathcal{A}\left([(Q_1\bar Q_3)_{1_c}(Q_2\bar Q_4)_{1_c}]_{1_c}\otimes |\psi_1\rangle\right).
\end{aligned}
\end{equation}
 Here, $Q_1\bar Q_3$ and $Q_2\bar Q_4$ form two subclusters in $|\psi_1\rangle$, while $Q_1\bar Q_4$ and $Q_2\bar Q_3$ form two subclusters in $|\psi_2\rangle$. The two subclusters are widely separated, therefore we have $\langle\psi_1|\psi_2\rangle\approx 0.$ We decompose the wave function in Eq.~\eqref{eq:wf_mt} in the orthogonal color basis as  
 \begin{eqnarray}
       && |\Psi\rangle=\chi_{1_c\otimes1_c}\otimes(|\psi_1\rangle+\frac{1}{3}|\psi_2\rangle)+\frac{2\sqrt{2}}{3}\chi_{8_c\otimes8_c}\otimes|\psi_2\rangle\\
     &&=\frac{1}{\sqrt{3}}\chi_{\bar3_c\otimes3_c}\otimes(|\psi_1\rangle-|\psi_2\rangle)  +\frac{\sqrt{2}}{\sqrt{3}}\chi_{6_c\otimes\bar6_c}\otimes(|\psi_1\rangle+|\psi_2\rangle)\nonumber,
 \end{eqnarray}
with
\begin{eqnarray}
\chi_{1_{c}\otimes1_{c}}&=[(Q_{1}\bar{Q}_{3})_{1_{c}}(Q_{2}\bar{Q}_{4})_{1_{c}}]_{1_{c}}\,,\\
\chi_{8_{c}\otimes8_{c}}&=[(Q_{1}\bar{Q}_{3})_{8_{c}}(Q_{2}\bar{Q}_{4})_{8_{c}}]_{1_{c}}\,,\\
\chi_{\bar{3}_{c}\otimes3_{c}}&=[(Q_{1}Q_{2})_{\bar{3}_{c}}(\bar{Q}_{3}\bar{Q}_{4})_{3_{c}}]_{1_{c}}\,,\\
\chi_{6_{c}\otimes\bar{6}_{c}}&=[(Q_{1}Q_{2})_{6_{c}}(\bar{Q}_{3}\bar{Q}_{4})_{\bar{6}_{c}}]_{1_{c}}\,.
\end{eqnarray}
For an ideal loose molecular state, we obtain $P_{1\otimes1}:P_{8\otimes8}=5:4$ and $P_{\bar3\otimes3}:P_{6\otimes\bar6}=1:2$. One can see that even for an ideal molecular state, the proportions of $\chi_{8_{c}\otimes8_c}$ and  $\chi_{1_{c}\otimes1_{c}}$ are comparable. Therefore,  discerning between molecular and compact states solely via the proportions of color components could be misleading in the systems with identical particles.}

Meanwhile, the naive definition of the rms radius could be misleading when the antisymmetric wave function is required for the identical quarks. For instance, when the mesons $(c\bar q)$ and $(s\bar q)$ form a molecular state, the wave function satisfying the Pauli principle is $|\psi_{\mathcal{A}}\rangle=|(c \bar{q}_1)(s \bar{q}_2)\rangle-|(c \bar{q}_2)(s \bar{q}_1)\rangle$. One can see that each antiquark $\bar q$ belongs to both mesons simultaneously. Therefore, neither $\langle\psi_{\mathcal{A}}|r_{c \bar{q}}^2| \psi_{\mathcal{A}}\rangle$ nor $\langle\psi_{\mathcal{A}}|r_{s \bar{q}}^2| \psi_{\mathcal{A}}\rangle$ can reflect the size of the constituent mesons. 
	
	In Ref.~\cite{Chen:2023syh}, we proposed a new approach to calculate the rms radii in the $ Qs\bar q\bar q $ system, which eliminates the ambiguity arising from the antisymmetrization of identical particles $ \bar q\bar q $. In the fully heavy tetraquark system, there exist two pairs of identical particles. Here we further extend the definitions of rms radii for the $ QQ\bar Q\bar Q $ system. We uniquely decompose the antisymmetric wave function as
	
	\begin{equation}\label{eq:wf_decompose}
		\begin{aligned}
			\Psi^J(\theta)=&\sum_{s_1\geq s_2}\left([(Q_1\bar Q_3)^{s_1}_{1_c}(Q_2\bar Q_4)^{s_2}_{1_c}]^{J}_{1_c}\otimes|\psi_1^{s_1s_2}(\theta)\rangle\right.\\
			&+[(Q_1\bar Q_3)^{s_2}_{1_c}(Q_2\bar Q_4)^{s_1}_{1_c}]^{J}_{1_c}\otimes|\psi_2^{s_1s_2}(\theta)\rangle\\
			&+[(Q_1\bar Q_4)^{s_1}_{1_c}(Q_2\bar Q_3)^{s_2}_{1_c}]^{J}_{1_c}\otimes|\psi_3^{s_1s_2}(\theta)\rangle\\
			&+\left.[(Q_1\bar Q_4)^{s_2}_{1_c}(Q_2\bar Q_3)^{s_1}_{1_c}]^{J}_{1_c}\otimes|\psi_4^{s_1s_2}(\theta)\rangle\right)\\
			=&\mathcal{A}\sum_{s_1\geq s_2}[(Q_1\bar Q_3)^{s_1}_{1_c}(Q_2\bar Q_4)^{s_2}_{1_c}]^{J}_{1_c}\otimes|\psi_1^{s_1s_2}(\theta)\rangle,
		\end{aligned}
	\end{equation}
	where $ s_1, s_2 $ sum over spin configurations with total angular momentum $ J $. We denote the non-antisymmetric component of the wave function as
	\begin{equation}
    \label{eq:nAwf}
		|\Psi_{\mathrm{nA}}^J(\theta)\rangle = \sum_{s_1\geq s_2}[(Q_1\bar Q_3)^{s_1}_{1_c}(Q_2\bar Q_4)^{s_2}_{1_c}]^{J}_{1_c}\otimes|\psi_1^{s_1s_2}(\theta)\rangle.
	\end{equation}
	where $ (Q_1\bar Q_3) $ and $ (Q_2\bar Q_4) $ form color singlets. Instead of using the complete wave function $ \Psi^J(\theta) $, we use $ |\Psi_{\mathrm{nA}}^J(\theta)\rangle $ to define the rms radius:
	\begin{equation}\label{eq:rmsr}
		r^{\mathrm{rms}}_{ij}\equiv \mathrm{Re}\left[\sqrt{\frac{\langle\Psi_{\mathrm{nA}}^J(\theta) | r_{ij}^2 e^{2i\theta}|\Psi_{\mathrm{nA}}^J(\theta)\rangle}{\langle\Psi_{\mathrm{nA}}^J(\theta) | \Psi_{\mathrm{nA}}^J(\theta)\rangle}}\right].
	\end{equation}
	\clabel[q1]{This definition discards the contribution of the exchange terms resulting from the antisymmetrization, which may play an important role in the numerical results of the rms radii, especially for compact tetraquark states. However, our primary interest lies in the general clustering behavior of the tetraquark states rather than in specific numerical results. The current definition of the rms radius is useful for investigating the internal spatial structures of the tetraquark states. } For example, if the resulting state is a scattering state or a hadronic molecule of $ \eta_c J/\psi $, $ r_{c_1\bar c_3} $ and $ r_{c_2\bar c_4} $ are respectively expected to be the sizes of $ J/\psi $ and $ \eta_c $, and much smaller than the other rms radii. On the other hand, if the resulting state is a compact tetraquark state, all rms radii in the four-body system should be of the same order. \change{One can find detailed examples for comparing two definitions of the rms radius in Appendix~\ref{app:rms}.} However, it should be noted that for a hadronic molecule composed of two mesons with the same quantum numbers but different radial excitation, for example $ \psi(3S)J/\psi $, the current definition cannot eliminate the ambiguity arising from the antisymmetrization. As a result, neither $ r_{c_1\bar c_3} $ nor $ r_{c_2\bar c_4} $ reflects the size of $ \psi(3S) $ or $ J/\psi $; instead, they represent the average of the sizes of the two mesons.
	
	It should be emphasized that the inner products in the CSM are defined using the c-product~\cite{ROMO1968617}, 
	\begin{equation}
		\langle\phi_n \mid \phi_m\rangle\equiv\int \phi_n(r)\phi_{m}(r)d^3r,
	\end{equation}
	where the square of the wave function rather than the square of its magnitude is used. The rms radius calculated by the c-product is generally not real, but its real part can still reflect the internal quark clustering behavior if the resonant state is not too broad, as discussed in Ref.~\cite{homma1997matrix}.
	
	\section{Results and Discussions}\label{sec:result}
	We investigate the S-wave fully charmed $ cc\bar c\bar c $ and fully bottomed $ bb\bar b\bar b $ tetraquark systems with all possible quantum numbers, including $ J^{PC}=0^{++},1^{+-},2^{++},0^{+-},1^{++},2^{+-} $. It should be stressed that the S-wave ground state diquarkonium thresholds exist only in the $0^{++},1^{+-},2^{++}$ systems, namely $ \eta_c\eta_c$ and $\eta_b\eta_b $ in the $ 0^{++} $ systems, $ \eta_cJ/\psi$ and $\eta_b\Upsilon $ in the $ 1^{+-} $ systems, $ J/\psi J/\psi$ and $\Upsilon\Upsilon $ in the $ 2^{++} $ systems. In the following discussions, the $0^{++},1^{+-},2^{++}$ systems are referred to as normal C-parity systems, whereas the $0^{+-},1^{++},2^{+-}$ systems are referred to as exotic C-parity systems. For convenience, we label the $ QQ\bar Q\bar Q $ tetraquark states obtained in the calculations as $ T_{4Q,J(C)}(M) $, where $ M $ is the mass of the state.
	
	\subsection{Fully charmed tetraquark}
	\subsubsection{States with normal C-parity}
\begin{figure*}[tbp]
		\centering
		\includegraphics[width=1\linewidth]{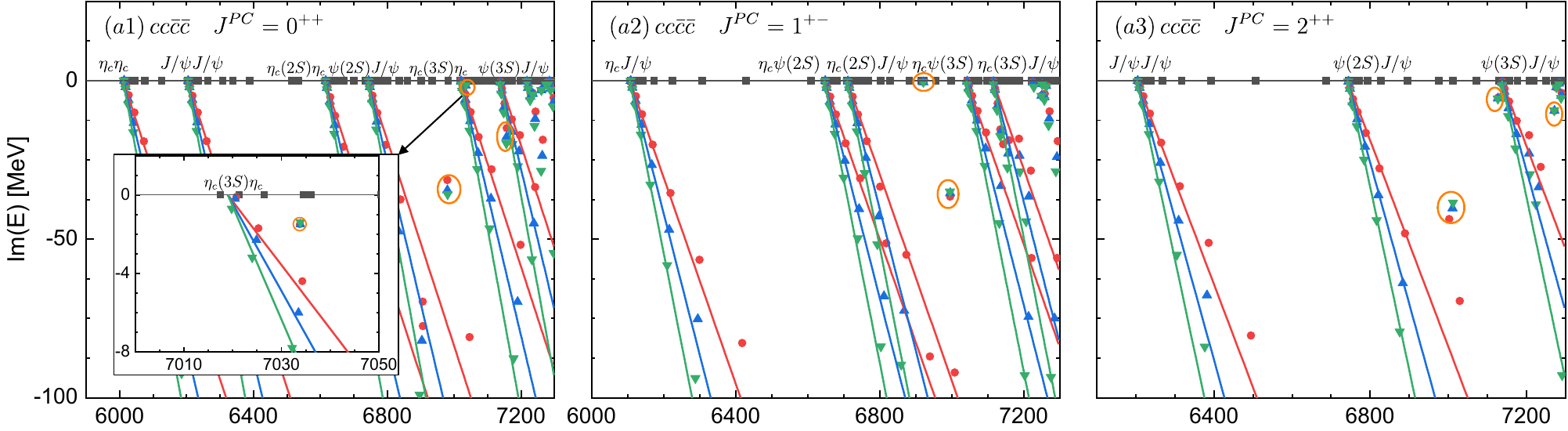}
		\includegraphics[width=1\linewidth]{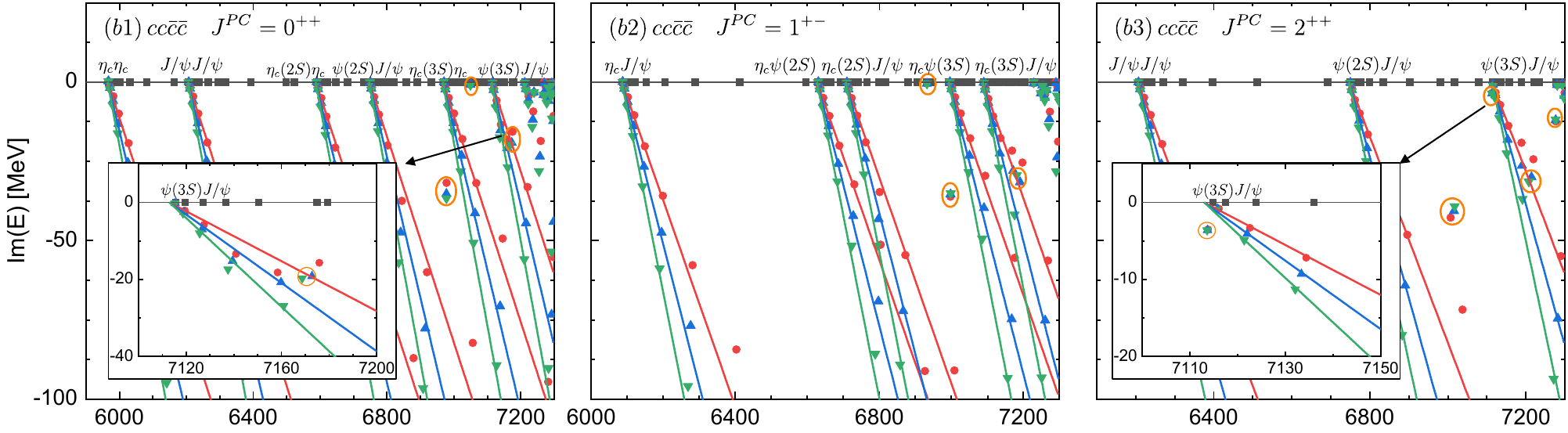}
		\includegraphics[width=1\linewidth]{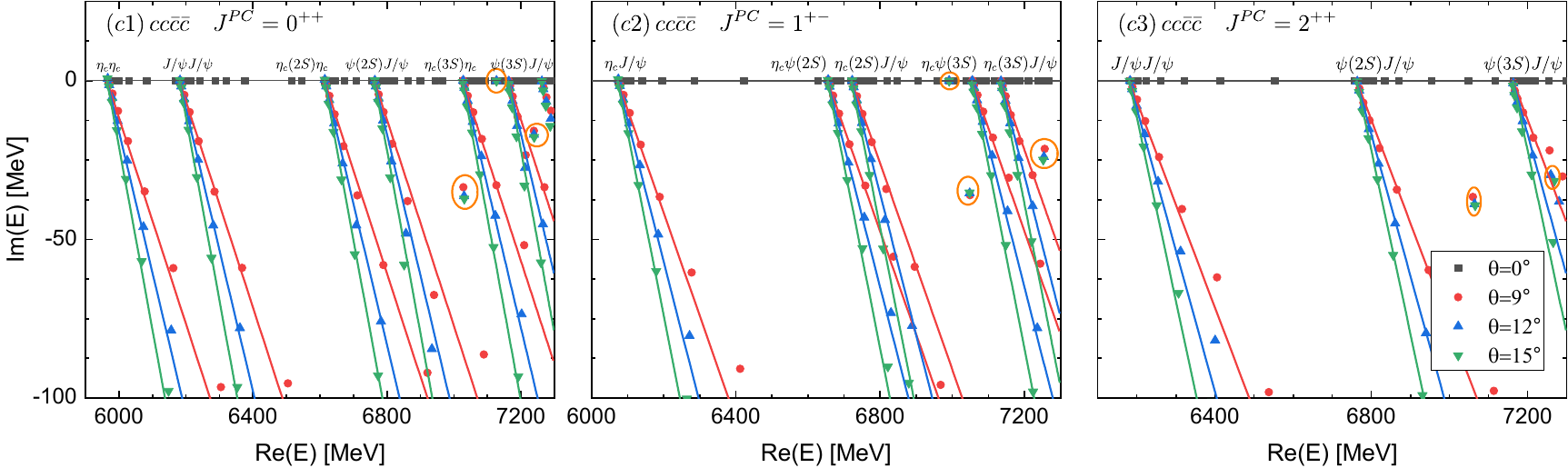}
		\caption{The complex energy eigenvalues of the  $cc\bar c\bar c$ states with normal C-parity in the (a) AL1, (b) AP1 and (c) BGS potential with varying $\theta$ in the CSM. The solid lines represent the continuum lines rotating along $\operatorname{Arg}(E)=-2 \theta$. The resonances do not shift as $\theta$ changes and are highlighted by the orange circles.}
		\label{fig:charm+}
	\end{figure*}

	With the CSM, the complex eigenenergies of the  $ 0^{++},1^{+-},2^{++} $ $ cc\bar c\bar c $ systems obtained from three different quark potential models are shown in Fig.~\ref{fig:charm+}. We choose varying complex scaling angles $ \theta $ to distinguish resonant states from scattering states. All of the states are above the lowest diquarkonium threshold, so no bound state is obtained. The diquarkonium scattering states rotate along the continuum lines starting from the threshold energies. Moreover, we obtain a series of resonant states whose complex energies are summarized in Table~\ref{tab:charm+_energies}. For comparison, we also list the results in Ref.~\cite{Wang:2022yes}, where the authors used the BGS potential for the $ cc\bar c\bar c $ systems.
	
	Qualitatively, the $ cc\bar c\bar c $ resonances obtained from three different quark potential models are in accordance with each other. Most of the resonances exist in all three models. For a specific resonant state, its width remains consistent across different models, while the mass in the BGS potential is approximately $ 50 $-$ 100 $ $\mathrm{MeV}$ larger than those in the AL1 and AP1 potentials. These differences are expected considering that the discrepancies of the predictions of the heavy quarkonium mass spectra from various potentials are up to tens of $\mathrm{MeV}$.
	
	The tetraquark resonant states with different quantum numbers $ J^{PC} $ exhibit a similar pattern. A lower resonant state with mass $ M\approx7000\,\mathrm{MeV} $ and width $ \Gamma\approx 75\,\mathrm{MeV}$, and a higher resonant state with mass $ M\approx7200\,\mathrm{MeV} $ and width $ \Gamma\approx 50\,\mathrm{MeV}$ are obtained in the $ 0^{++}, 1^{+-}, 2^{++} $ systems. The lower $ 0^{++} $ state can decay into the $ \eta_c\eta_c , J/\psi J/\psi, \eta_c(2S)\eta_c $ and $ \psi(2S)J/\psi $ channels. Additionally, the higher $ 0^{++} $ state can decay into the $ \eta_c(3S)\eta_c $ and $ \psi(3S)J/\psi $ channels. The lower $ 2^{++} $ state can decay into the $ J/\psi J/\psi $ and $ \psi(2S)J/\psi $ channels. Additionally, the higher $ 2^{++} $ state can decay into the $ \psi(3S)J/\psi $ channel. Considering that the quark potential models have errors up to tens of $ \mathrm{MeV} $ and that we have neglected the widths of the quarkonia in our calculations, the lower $ 0^{++} $ and $ 2^{++} $ states may serve as the candidates for the experimentally observed $ X(6900) $ state, while the higher $ 0^{++} $ and $ 2^{++} $ states may serve as the candidates for the experimentally observed $ X(7200) $ state. On the other hand, the lower $ 1^{+-} $ state can decay into the $ \eta_cJ/\psi, \eta_c(2S)J/\psi $ and $ \eta_c\psi(2S) $ channels. Additionally, the higher $1^{+-}$ state can decay into the $ \eta_c\psi(3S) $ and $ \eta_c(3S)J/\psi $ channels. The $ 1^{+-} $ states are not the candidates for $ X(6900) $ or $ X(7200) $ because they cannot decay into either the $ J/\psi J/\psi $ or $ \psi(2S)J/\psi $ channels. These states can be searched for in future experiments.
	
	Moreover, we observe several narrow resonant states with different quantum numbers. These states are found in the mass region  $ (6.92, 7.30)\,\mathrm{GeV}$. 	These narrow resonances can be searched for by experiments in the corresponding diquarkonium decay channels. However, we do not observe any signal for resonance in the mass region $ (6.2, 6.6)\,\mathrm{GeV} $, namely no candidate for $ X(6400) $ or $ X(6600) $ is found. 
	
	Comparing the results of resonant states obtained from the BGS potential with those of Ref.~\cite{Wang:2022yes}, our calculations can reproduce the previous results well within numerical uncertainty. In addition, we obtain three extra narrow resonant states in the BGS potential. The reason for these discrepancies might be that a set of complete color-spin basis is used in our calculations while some basis functions are neglected in Ref.~\cite{Wang:2022yes}. For example, in the $ 0^{++} $ system, the diquarkonium color configuration $ [(Q_1\bar Q_3)_{8_c}(Q_2\bar{Q}_4)_{8_c}]_{1_c} $ is not included in the previous calculations. 
	These missing basis functions turn out to be crucial for the existence of the extra resonant states. 
	
	As mentioned above, the fully charmed tetraquark resonant states obtained from different models qualitatively agree with each other. Therefore, we only choose the results in the AP1 potential to analyze their inner structures. The proportions of the color configurations $ \chi_{\bar 3_c\otimes 3_c} $ and $ \chi_{6_c\otimes \bar6_c} $ in the wave functions as well as the rms radii of the fully charmed resonant states are listed in Table~\ref{tab:charm+_structure}. For most resonant states, their rms radii are approximately of the same size and less than $ 1 $ fm, indicating that they are compact tetraquark states. However, for the $ X(7200) $ candidate states $ T_{4c,0(+)}(7173), T_{4c,2(+)}(7214) $ and the state $ T_{4c,1(-)}(7191) $ , the $ r^{\rm rms}_{c_1\bar c_3} $ and $ r^{\rm rms}_{c_2\bar c_4} $ are much smaller than the other radii. Compared with the rms radii of the quarkonia listed in Table~\ref{tab:meson}, we observe that both $ r^{\rm rms}_{c_1\bar c_3} $ and $ r^{\rm rms}_{c_2\bar c_4} $ are larger than the rms radii of $\eta_c$ and $ J/\psi $, and smaller than those of $\eta_c(3S) $ and $ \psi(3S) $, while  $ r^{\rm rms}_{c_1\bar c_4}, r^{\rm rms}_{c_2\bar c_3}, r^{\rm rms}_{c_1c_2} $ and $ r^{\rm rms}_{\bar c_3\bar c_4} $ are much larger than the rms radii of all mesons. These results suggest that these three states might have a molecular configuration. It should be noted that these resonant states have relatively large widths and are located close to the continuum line of the scattering states, therefore the results of their rms radii are less numerically accurate in the CSM and should be considered as qualitative estimates.
	\begin{table}[htbp]
		\centering
		\caption{The complex energies $ E=M-i\Gamma/2 $ (in $\mathrm{MeV}$) of the $ cc\bar c\bar c $ resonant states with normal C-parity from various potential models. The last column lists the results in Ref.~\cite{Wang:2022yes}. The  ``?"  indicates the potential existence of resonant states, which blend into the continuum lines of scattering states and cannot be obtained accurately in the present calculations. The ``-" suggests that no corresponding resonance is obtained.   }
		\label{tab:charm+_energies}
		\begin{tabular}{ccccc}
			\hline\hline
			$ J^{PC} $& AL1 & AP1& BGS & BGS, Wang \textit{et al.}\\
			\hline
			$ 0^{++} $&$ 6980-35i $&  $ 6978-36i	 $&$ 7030-36i $&$ 7035-39i $\\
			&$ 7034-1i $&$ 7049-1i $&$ 7127-0.1i $&-\\
			&$ 7156-20i $&$ 7173-20i $&$ 7239-17i $&$ 7202-30i $\\
			
			$ 1^{+-} $&$ 6921-0.5i $&$ 6932-0.5i $&$ 6991-0.1i $& - \\
			&$ 6995-35i $&$ 6998-35i $&$ 7048-35i $&$ 7050-35i $\\
			&?&$ 7191-32i $&$ 7254-24i $&$ 7273-25i $\\
			$ 2^{++} $&$ 7013-38i $&$ 7017-39i$&$ 7066-39i$&$ 7068-42i $\\
			&$ 7127-6i $&$ 7114-4i $&-&-\\
			&?&$ 7214-30i $&$ 7268-32i $&$ 7281-46i $\\
			& $ 7272-9i $ &$ 7276-12i $&$ 7337-8i $&-\\
			\hline\hline
		\end{tabular}
	\end{table}
	
	\begin{table*}
		\centering
		\caption{The proportions of different color configurations and the rms radii (in fm) of the $ cc\bar c\bar c $ resonant states with normal C-parity in the AP1 potential. The last column shows the spatial configurations of the states, where C. and M. represent the compact tetraquark and molecular configurations, respectively. }
		\label{tab:charm+_structure}
		\begin{tabular*}{\hsize}{@{}@{\extracolsep{\fill}}ccccccccc@{}}
			\hline\hline
			$ J^{PC} $&$  M-i\Gamma/2 $ &$ \chi_{\bar 3_c\otimes 3_c} $& $ \chi_{6_c\otimes\bar6_c} $& $ r_{c_1\bar{c}_3}^{\mathrm{rms}} $&$ r_{c_2\bar{c}_4}^{\mathrm{rms}} $&$ r_{c_1\bar{c}_4}^{\mathrm{rms}} $=$ r_{c_2\bar{c}_3}^{\mathrm{rms}} $&$ r_{c_1c_2}^{\mathrm{rms}} $=$ r_{\bar{c}_3\bar{c}_4}^{\mathrm{rms}} $& Configurations\\
			\hline
			
			$ 0^{++} $&$ 6978-36i $&$ 86\% $&$14\%$&$0.81$&$0.81$&$0.86$&$0.66$&C.\\
			&$ 7049-1i $&$37\%$&$63\%$&$0.70$&$0.70$&$0.82$&$0.75$&C.\\
			&$ 7173-20i $&$46\%$&$54\%$&$0.89$&$0.89$&$2.31$&$2.28$&M.\\
			$ 1^{+-} $&$ 6932-0.5i $&$65\%$&$35\%$&$0.66$&$0.66$&$0.73$&$0.63$&C.\\
			&$ 6998-35i $&$88\%$&$12\%$&$0.79$&$0.80$&$0.77$&$0.59$&C.\\
			&$ 7191-32i $&$44\%$&$56\%$&$0.71$&$1.08$&$2.09$&$2.08$&M.\\
			$ 2^{++} $&$ 7017-39i $&$90\%$&$10\%$&$0.79$&$0.79$&$0.71$&$0.56$&C.\\
			&$ 7114-4i $&$69\%$&$31\%$&$0.92$&$0.92$&$0.65$&$0.55$&C.\\
			&$ 7214-30i $&$57\%$&$43\%$&$0.92$&$0.92$&$1.93$&$1.88$&M.\\
			&$ 7276-12i $&$73\%$&$27\%$&$0.86$&$0.86$&$1.04$&$0.93$&C.\\
			\hline\hline
			
		\end{tabular*}
	\end{table*}
	
	\subsubsection{States with exotic C-parity}
\begin{figure*}[tbp]
		\centering
		\includegraphics[width=1\linewidth]{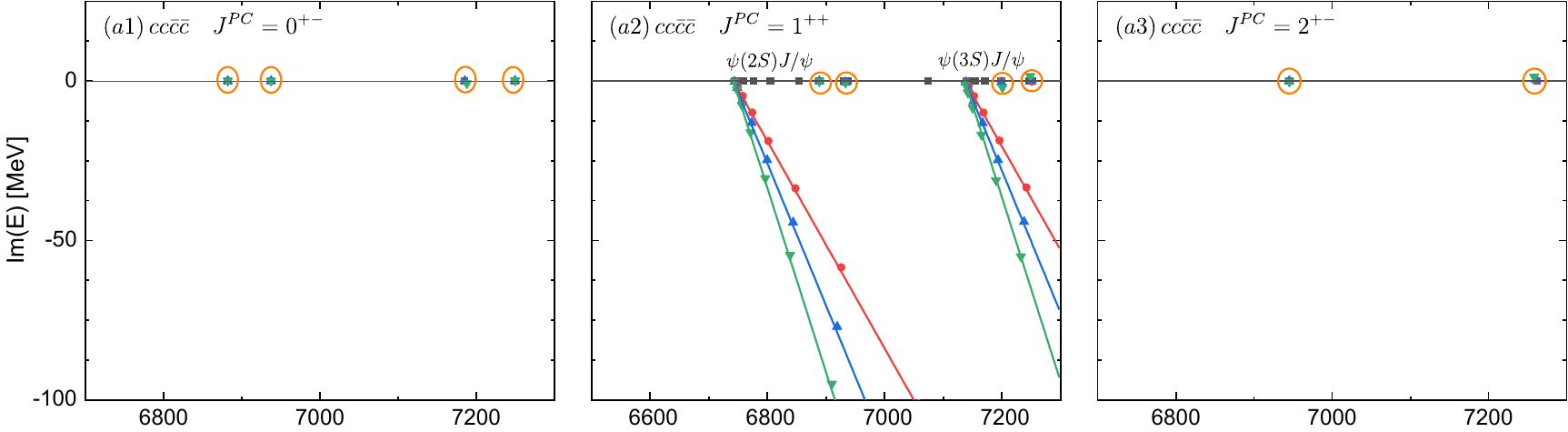}
		\includegraphics[width=1\linewidth]{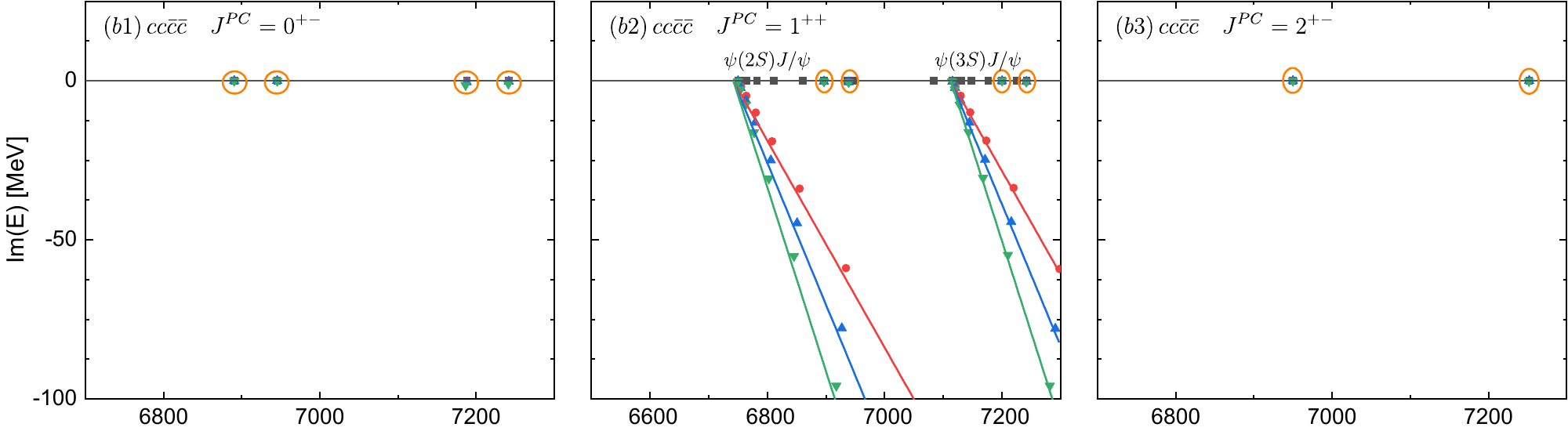}
		\includegraphics[width=1\linewidth]{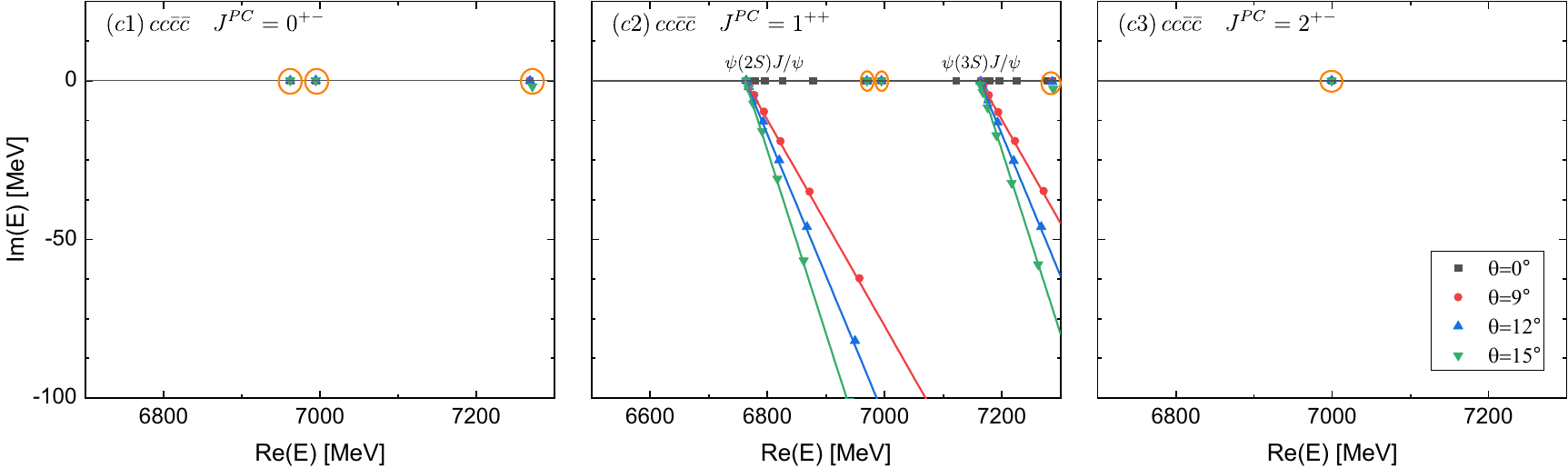}
		\caption{The complex energy eigenvalues of the  $cc\bar c\bar c$ states with exotic C-parity in the (a) AL1, (b) AP1 and (c) BGS potential with varying $\theta$ in the CSM. The solid lines represent the continuum lines rotating along $\operatorname{Arg}(E)=-2 \theta$. The zero-width states and resonances do not shift as $\theta$ changes and are highlighted by the orange circles.}
		\label{fig:cccc-}
	\end{figure*}

	The complex eigenenergies of the fully charmed tetraquark systems with exotic C-parity $ (J^{PC}=0^{+-}, 1^{++}, 2^{+-}) $ obtained from three different quark potential models are shown in Fig.~\ref{fig:cccc-}. We obtain a series of resonant and zero-width states, whose energies are summarized in Table~\ref{tab:charm-_energies}. Similar to the systems with normal C-parity, the states with exotic C-parity obtained from different models qualitatively agree with each other. The masses of a specific state in the AL1 and AP1 potentials are nearly the same, while the mass in the BGS potential is around  $ 50 $-$ 100 $ $\mathrm{MeV}$ larger than the former ones. In the following we solely focus on the results in the AP1 potential. The proportions of the color configurations and the rms radii of the resonant and zero-width states are listed in Table~\ref{tab:charm-_structure}. The different rms radii of these states are approximately the same and less than 1 fm, indicating that all of these states have compact tetraquark configuration.
	
	For the $ 1^{++} $ $ cc\bar c\bar c $ system,
	we obtain a series of $ \psi(2S)J/\psi $ and $ \psi(3S)J/\psi $ diquarkonium scattering states as well as four extremely narrow resonant states, whose two-body decay widths are less than $ 1 $ $ \mathrm{MeV}$. The states $ T_{4c,1(+)}(6939) $ and $ T_{4c,1(+)}(7242) $ have predominant color configuration $ \chi_{\bar 3_c\otimes 3_c} $.
    For the states $ T_{4c,1(+)}(6896) $ and $ T_{4c,1(+)}(7200) $, the proportions of $ \chi_{\bar 3_c\otimes 3_c} $ and $ \chi_{6_c\otimes \bar6_c} $ are around $ \frac{2}{3} $ and $ \frac{1}{3} $, respectively. 
    The states $ T_{4c,1(+)}(7200) $ and  $ T_{4c,1(+)}(7242) $ are the radial excitation of $ T_{4c,1(+)}(6896) $ and $ T_{4c,1(+)}(6939) $, respectively. All of the four states can decay into the $ \psi(2S)J/\psi $ channel, but the decay widths are very small. In Sec.~\ref{subsec:small_width}, we discuss the reasons for the small widths in detail.
	
	For the $ 0^{+-} $ $ cc\bar c\bar c $ system, there do no exist any S-wave diquarkonium thresholds. Therefore no meson-meson scattering state is observed. All of the states in the system are identified as zero-width states, which lie on the real axis and are not changed by the complex scaling. According to the proportions of color configurations listed in Table~\ref{tab:charm-_structure}, the four $ 0^{+-} $ zero-width states can be clearly arranged into two doublets, $\left\{ T_{4c,0(-)}(6890), T_{4c,0(-)}(6945)\right\}$ and $ \left\{T_{4c,0(-)}(7188), T_{4c,0(-)}(7242)\right\} $. The color configurations of the two states inside a doublet are orthogonal to each other. The lower state is dominated by the $ \chi_{6_c\otimes\bar6_c} $ component, while the higher state is dominated by the $ \chi_{\bar3_c\otimes3_c} $ component.  In the fully heavy tetraquark system, it is known that the color magnetic term is suppressed by the heavy quark mass, and the dominant color electric interactions between two (anti)quarks are attractive in $ [Q_1Q_2]_{\bar 3_c} $ and $ [\bar Q_3\bar Q_4]_{3_c} $ configurations but repulsive in $ [Q_1Q_2]_{6_c} $ and $ [\bar Q_3\bar Q_4]_{\bar 6_c} $ configurations. Besides, the color electric term also provides an attractive interaction between the $ 6_c $ diquark and $ \bar 6_c $ antidiquark, which is much stronger than the one between $ \bar 3_c $ diquark and $ 3_c $ antidiquark due to the color SU(3) algebra~\cite{Wang:2019rdo,Deng:2020iqw,Ader:1981db}. The fact that the lower state has the dominant $ \chi_{6_c\otimes\bar6_c} $ configuration suggests that the strong attraction between two color sextet clusters prevails over the repulsion within the (anti)diquark and contributes to the formation of a deeper state than the $ \chi_{\bar3_c\otimes3_c} $ dominant one, which is consistent with the conclusion in Refs.~\cite{Wang:2019rdo,Ader:1981db}. As a result of the interaction mechanism, the rms radii $ r^{\rm rms}_{c_1c_2} $ and $ r^{\rm rms}_{\bar c_3\bar c_4} $, which characterize the sizes of the diquark and antidiquark, take larger values in the $ \chi_{6_c\otimes \bar6_c} $ dominant state than in the $ \chi_{\bar3_c\otimes 3_c} $ dominant state. On the other hand, the rms radii $ r^{\rm rms}_{c_1\bar c_3} $, $ r^{\rm rms}_{c_2\bar c_4} $, $ r^{\rm rms}_{c_1\bar c_4} $ and $ r^{\rm rms}_{c_2\bar c_3} $, which characterize the distance between diquark and antidiquark, take smaller values in the $ \chi_{6_c\otimes \bar6_c} $ dominant state than in the $ \chi_{\bar 3_c\otimes 3_c} $ dominant state. \clabel[q2]{It is also worth mentioning that the four rms radii $r^{\rm rms}_{c_i\bar c_j}$ equal to each other in the $0^{+-}$ and $2^{+-}$ systems. The reason is that these states consist only of S-wave diquark-antidiquark configuration, as illustrated in Appendix.~\ref{app:basis}, and their spatial wave function is symmetric under the particle exchange $c_1\leftrightarrow c_2$ or $\bar c_3\leftrightarrow\bar c_4$.} The higher doublet states $ \left\{T_{4c,0(-)}(7188), T_{4c,0(-)}(7242)\right\} $ have larger rms radii than the lower doublet states $\left\{ T_{4c,0(-)}(6890), T_{4c,0(-)}(6945)\right\}$ and can be viewed as the radial excitation of the latter.
	
	Similar to the $ 0^{+-} $ system, S-wave diquarkonium threshold does not exist in the $ 2^{+-} $ $ cc\bar c\bar c $ system, and all of the states in the system are identified as zero-width states. Due to the restriction of antisymmetrization of identical particles, the only allowed color configuration for these states is $ \chi_{\bar 3_c\otimes 3_c} $. The state $ T_{4c,2(-)}(7252) $ is the radial excitation of the ground state $ T_{4c,2(-)}(6950) $. 
	
	It should be noted that although S-wave diquarkonium threshold does not exist in the $ 0^{+-} $ and $ 2^{+-} $ $ cc\bar c\bar c $ systems, diquarkonium thresholds with higher orbital angular momentum do exist in these systems. For example, the P-wave $ J/\psi\chi_{c1} $ state and the S-wave $ h_c\chi_{c1} $ state can form the $ 0^{+-} $ or $ 2^{+-} $ $ cc\bar c\bar c $ system, while the S-wave $ \eta_c\psi_2 $ state and the P-wave  $ J/\psi\chi_{c2} $ can form the $ 2^{+-} $ $ cc\bar c\bar c $ system. These scattering states have the same quantum numbers as the zero-width states obtained in the present calculations. The coupling between them may alter the positions of the zero-width states. Considering the effect of this coupling is beyond the scope of this work. However, if one assumes the coupling effect is small and treats it as perturbation, the positions of the states should not change by much. The zero-width states may obtain nonzero widths and transform into resonant states, which can decay into the diquarkonium channels with lower energies. All of these states lie above the ground state diquarkonium threshold $ J/\psi\chi_{c1} $,  whose theoretical energy in the AP1 potential is $ 6593\,\mathrm{MeV} $. Therefore, they may be searched for in the P-wave $ J/\psi\chi_{c1} $ decay channel in the experiment.
	
	\begin{table}[htbp]
		\centering
		\caption{The complex energies $ E=M-i\Gamma/2 $ (in $\mathrm{MeV}$) of the $ cc\bar c\bar c $ resonant and zero-width states with exotic C-parity from various potential models.}
		\label{tab:charm-_energies}
		\begin{tabular*}{\hsize}{@{}@{\extracolsep{\fill}}cccc@{}}
			\hline\hline
			$ J^{PC} $& AL1 & AP1& BGS\\
			\hline
			$ 0^{+-} $&$ 6882 $&$ 6890 $&$ 6962 $\\
			&$ 6938 $&$ 6945 $&$ 6995 $\\
			&$ 7185 $&$ 7188 $&$ 7268 $\\
			&$ 7249 $&$ 7242 $&$ 7316 $\\
			$ 1^{++} $&$ 6889-0.1i $&$ 6896-0.2i $&$ 6970-0.1i $\\
			&$ 6933-0.4i $&$ 6939-0.4i $&$ 6995-0.1i $\\
			&$ 7200-0.4i $&$ 7200-0.1i $&$ 7286-0.4i $\\
			&$ 7251-0.1i $&$ 7242-0.3i $&$ 7318-0.2i $\\
			$ 2^{+-} $&$ 6945 $&$ 6950 $&$ 6999 $\\
			&$ 7262 $&$  7252 $ &$ 7324 $\\
			\hline\hline
		\end{tabular*}
	\end{table}
	
	\begin{table*}
		\centering
		\caption{The proportions of different color configurations and the rms radii (in fm) of the $ cc\bar c\bar c $ resonant and zero-width states with exotic C-parity in the AP1 potential. The last column shows the spatial configurations of the states, where C. and M. represent the compact tetraquark and molecular configurations, respectively. }
		\label{tab:charm-_structure}
		\begin{tabular*}{\hsize}{@{}@{\extracolsep{\fill}}ccccccccc@{}}
			\hline\hline
			$ J^{PC} $&$  M-i\Gamma/2 $ & $ \chi_{\bar 3_c\otimes 3_c} $& $ \chi_{6_c\otimes\bar6_c} $& $ r_{c_1\bar{c}_3}^{\mathrm{rms}} $&$ r_{c_2\bar{c}_4}^{\mathrm{rms}} $&$ r_{c_1\bar{c}_4}^{\mathrm{rms}} $=$ r_{c_2\bar{c}_3}^{\mathrm{rms}} $&$ r_{c_1c_2}^{\mathrm{rms}} $=$ r_{\bar{c}_3\bar{c}_4}^{\mathrm{rms}} $&Configuration\\
			\hline
			
			$ 0^{+-} $&$ 6890 $&$ 36 \%$&$ 64 $\%&$ 0.62 $&$ 0.62 $&$ 0.62 $&$ 0.71 $&C.\\
			&$ 6945 $&$ 64 \%$&$ 36 \%$&$ 0.65 $&$ 0.65 $&$ 0.65 $&$ 0.71 $&C.\\
			&$ 7188 $&$ 18 \%$&$ 82 \%$&$ 0.80 $&$ 0.80 $&$ 0.80 $&$ 0.93 $&C.\\
			&$ 7242 $&$ 82 \%$&$ 18 \%$&$0.85$&$0.85$&$0.85$&$0.83$&C.\\
			$ 1^{++} $&$ 6896-0.2i$&$71\%$&$29\%$&$ 0.67$&$0.58$&$0.71$&$0.63$&C.\\
			&$ 6939-0.4i $&$96\%$&$4\%$&$0.60$&$0.68$&$0.65$&$0.68$&C.\\
			&$ 7200-0.1i $&$68\%$&$32\%$&$0.83$&$0.75$&$0.94$&$0.78$&C.\\
			&$ 7242-0.3i $&$98\%$&$2\%$&$0.82$&$0.88$&$0.85$&$0.81$&C.\\
			$ 2^{+-} $&$6950$&$100\%$&$0\%$&$0.65$&$0.65$&$0.65$&$0.69$&C.\\
			&$7252$&$100\%$&$0\%$&$0.86$&$0.86$&$0.86$&$0.81$&C.\\
			\hline\hline
			
		\end{tabular*}
	\end{table*}
	
	\subsection{Fully bottomed tetraquark}
\begin{figure*}[tbp]
		\centering
		\includegraphics[width=1\linewidth]{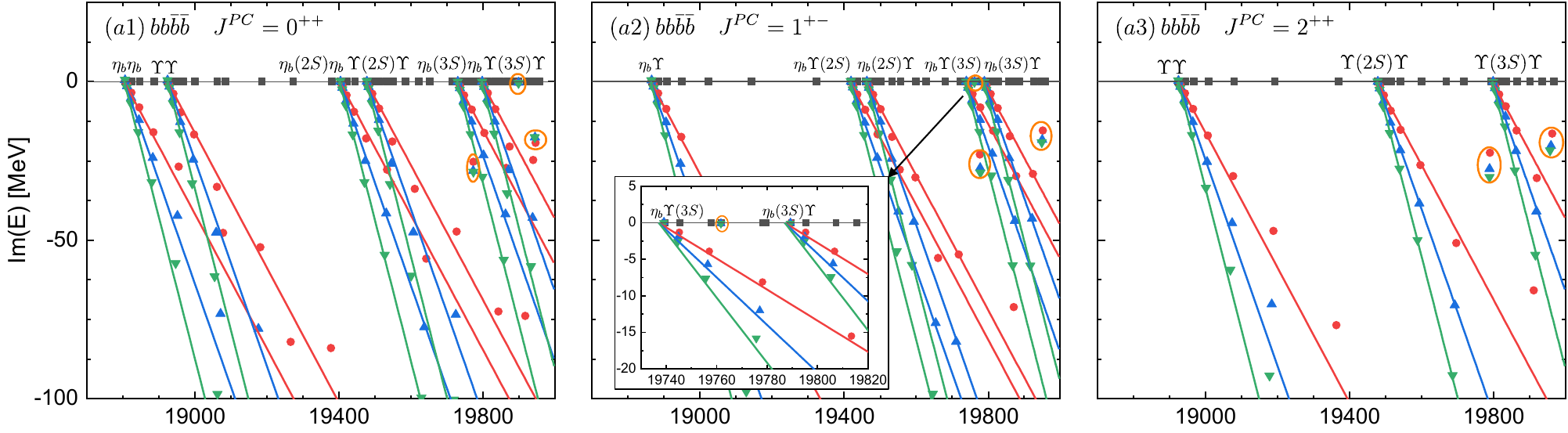}
		\includegraphics[width=1\linewidth]{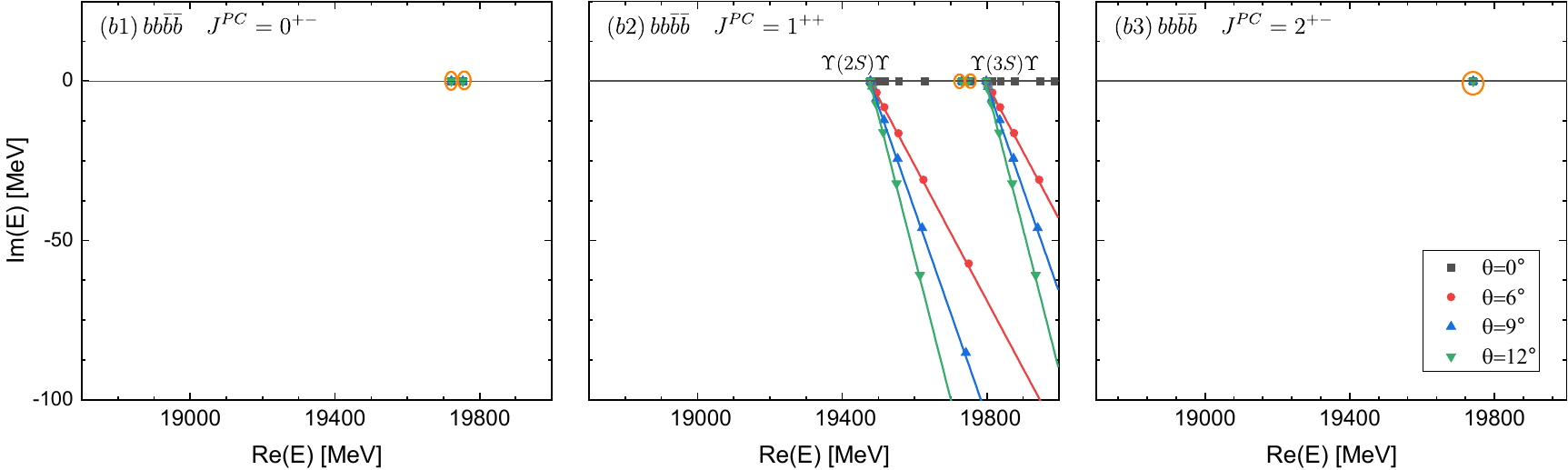}
		\caption{The complex energy eigenvalues of the  $bb\bar b\bar b$ states with (a) normal and (b) exotic C-parities in the AP1 potential with varying $\theta$ in the CSM. The solid lines represent the continuum lines rotating along $\operatorname{Arg}(E)=-2 \theta$. The zero-width states and resonances do not shift as $\theta$ changes and are highlighted by the orange circles.}
		\label{fig:bottom}
	\end{figure*}

	In the $ cc\bar c\bar c $  system, we observe that the results in different quark potential models qualitatively agree with each other. Therefore, we only choose the AP1 potential to investigate the $ bb\bar b\bar b $ system. The complex eigenenergies of the $ bb\bar b\bar b $ states with normal and exotic C-parities are shown in Fig.~\ref{fig:bottom}.
	
	Similar to the $ cc\bar c\bar c $ system, we obtain a series of resonant states with normal C-parity in the $ bb\bar b\bar b $ system. The complex energies, the proportions of different color configurations and the rms radii of these resonant states are listed in Table~\ref{tab:bottom+_structure}.
	The different rms radii of these states are around $ 0.5 $ fm, indicating that all of these states have compact tetraquark configuration. We obtain a lower resonant state with mass $ M\approx19780\,\mathrm {MeV}$ and width $\Gamma\approx60\,\mathrm {MeV}$, and a higher resonant state with mass $M\approx19950\,\mathrm {MeV}$ and width $\Gamma\approx40\,\mathrm{MeV}$ in the $0^{++}, 1^{+-}$ and $2^{++}$ systems. We also obtain two narrow resonant states $T_{4b,0(+)}(19898)$ and $T_{4b,1(-)}(19762)$. These states can decay strongly and be searched for in the corresponding diquarkonium decay channels in the experiment.

	For the $bb\bar b\bar b$ system with exotic C-parity, we obtain a series of resonant and zero-width states, whose complex energies, proportions of different color configurations and rms radii are listed in Table~\ref{tab:bottom-_structure}. For the $0^{+-}$ system, we obtain two zero-width states $T_{4b,0(-)}(19722)$ and $T_{4b,0(-)}(19754)$. The lower one is dominated by the $\chi_{\bar3_c\otimes 3_c}$ color configuration while the higher one is dominated by the $\chi_{6_c\otimes \bar6_c}$ color configuration. The mass hierarchy of the $\chi_{\bar3_c\otimes 3_c}$ dominated state and the $\chi_{6_c\otimes \bar6_c}$ dominated state is reversed compared to the $cc\bar c\bar c$ system. This suggests that in the $bb\bar b\bar b$ system, the interaction within the (anti)diquark plays a more important role than the interaction between diquark and antidiquark.  For the $1^{++}$ system, we obtain two extremely narrow resonant states $T_{4b,1(+)}(19731)$ and $T_{4b,1(+)}(19748)$, whose two-body decay widths are less than $0.1\,\mathrm{MeV}$.  
    For the $2^{++}$ system, a zero-width state $T_{4b,2(-)}(19741)$ is obtained. These states may couple with the P-wave diquarkonium thresholds. Due to the coupling effect, the $ 0^{+-} $ and $2^{+-}$ zero-width states may transform into resonant states, which can decay into the P-wave $ \Upsilon\chi_{b1} $ channel.

	\begin{table*}
		\centering
		\caption{The proportions of different color configurations and the rms radii (in fm) of the $ bb\bar b\bar b $ resonant states with normal C-parity in the AP1 potential. The last column shows the spatial configurations of the states, where C. and M. represent the compact tetraquark and molecular configurations, respectively. }
		\label{tab:bottom+_structure}
		\begin{tabular*}{\hsize}{@{}@{\extracolsep{\fill}}ccccccccc@{}}
			\hline\hline
			$ J^{PC} $&$  M-i\Gamma/2 $ & $ \chi_{\bar3_c\otimes3_c} $ &$ \chi_{6_c\otimes\bar6_c} $& $ r_{b_1\bar{b}_3}^{\mathrm{rms}} $& $ r_{b_2\bar{b}_4}^{\mathrm{rms}} $& $ r_{b_1\bar{b}_4}^{\mathrm{rms}}=r_{b_2\bar{b}_3}^{\mathrm{rms}} $&$ r_{b_1b_2}^{\mathrm{rms}} $=$ r_{\bar{b}_3\bar{b}_4}^{\mathrm{rms}} $&Configuration\\
			\hline
			
			$ 0^{++} $&$ 19773-28i $&$ 87\% $&$ 13\% $&$0.51$&$0.51$&$0.50$&$0.33$&C.\\
			&$ 19898-0.4i $&$ 36\% $& $ 64\% $&$0.43$&$0.43$&$0.50$&$0.47$&C.\\
			&$ 19944-17i $& $ 80\% $ & $ 20\% $&$0.58$&$0.58$&$0.56$&$0.38$&C.\\
			
			$ 1^{+-} $&$ 19762-0.1i $&$ 65\% $&$ 35\% $&$ 0.40$&$ 0.40 $& $ 0.44$&$0.38 $&C.\\
			&$ 19778-29i$&$ 87\% $&$ 13\% $&$0.51$&$0.51$&$0.52$&$0.36$&C.\\
			&$ 19948-19i $&$78\%$&$ 22\% $&$0.58$&$0.58$&$0.59$&$0.41$&C.\\
			$ 2^{++} $&$ 19788-30i $&$ 86\% $&$ 14\%$&$0.51$&$0.51$&$0.53$&$0.38$&C.\\
			&$ 19957-22i $&$ 76\% $&$ 24\% $&$0.59$&$0.59$&$0.62$&$0.45$&C.\\
			\hline\hline
		\end{tabular*}
	\end{table*}
	
	\begin{table*}
		\centering
		\caption{The proportions of different color configurations and the rms radii (in fm) of the $ bb\bar b\bar b $ resonant and zero-width states with exotic C-parity in the AP1 potential. The last column shows the spatial configurations of the states, where C. and M. represent the compact tetraquark and molecular configurations, respectively.}
		\label{tab:bottom-_structure}
		\begin{tabular*}{\hsize}{@{}@{\extracolsep{\fill}}ccccccccc@{}}
			\hline\hline
			$ J^{PC} $&$  M-i\Gamma/2 $ & $ \chi_{\bar3_c\otimes3_c} $ &$ \chi_{6_c\otimes\bar6_c} $& $ r_{b_1\bar{b}_3}^{\mathrm{rms}} $& $ r_{b_2\bar{b}_4}^{\mathrm{rms}} $& $ r_{b_1\bar{b}_4}^{\mathrm{rms}}=r_{b_2\bar{b}_3}^{\mathrm{rms}} $&$ r_{b_1b_2}^{\mathrm{rms}} $=$ r_{\bar{b}_3\bar{b}_4}^{\mathrm{rms}} $&Configuration\\
			\hline
			
			$ 0^{+-} $&$ 19722 $&$ 77\% $&$ 23\% $&$0.38$&$0.38$&$0.38$&$0.41$&C.\\
			&$ 19754 $&$ 23\% $& $ 77\% $&$0.39$&$0.39$&$0.39$&$0.45$&C.\\

			$ 1^{++} $&$ 19731 $&$ 96\% $&$ 4\%
 $&$0.41$&$0.36$&$0.39$&$0.41$&C.\\
			&$ 19748 $&$71\%$&$ 29\% $&$0.35$&$0.42$&$0.44$&$0.38$&C.\\
			$ 2^{+-} $&$ 19741 $&$ 100\% $&$ 0\% $&$0.39$&$0.39$&$0.39$&$0.42$&C.\\
			\hline\hline
		\end{tabular*}
	\end{table*}

	\subsection{Resonances with small widths}\label{subsec:small_width}
	In our calculations, we obtain several extremely narrow resonant states with quantum numbers $ J^P=1^+ $ in both the $ cc\bar c\bar c $ and $ bb\bar b\bar b $ systems. Intuitively, one would assume that these states can decay into the diquarkonium channels with the same quantum numbers and lower energies. However, the narrow widths of these states indicate that the two-body decay process is suppressed. The reasons for the suppression are given as follows.
	
	The two-body decay width of the tetraquark state is proportional to the modulus square of the $ T $-matrix element~\cite{Xiao:2019spy},
	\begin{equation}
		\Gamma_{F\rightarrow A+B}\propto\left|T_{F\rightarrow A+B}\right|^2=\left|\langle\psi_{AB}|\sum_{i<j=1}^4V_{ij}|\psi_{F}\rangle\right|^2,
	\end{equation}
	where $ F $ denotes the tetraquark state and $ A, B $ denote the final mesons. The potential $ V_{ij} $
	is given in Eq.~\eqref{eq:potential}, comprising the color magnetic term and spin-independent terms. In the fully heavy system, the color magnetic transition is suppressed by the heavy quark mass. The spin-independent terms contain the color factor $ \boldsymbol\lambda_i\cdot\boldsymbol\lambda_j $, whose matrix elements are listed in Appendix~\ref{app:color}. It can be seen that $ \sum_{i<j=1}^4\boldsymbol\lambda_i\cdot\boldsymbol\lambda_j $ is proportional to the identity operator and cannot induce transition between different color configurations. Color mixing via spin-independent terms can occur only when the coefficients of $ \boldsymbol\lambda_i\cdot\boldsymbol\lambda_j $ are different. Roughly speaking, the magnitudes of color mixing matrix elements depend on the differences between various coefficients. From Tables~\ref{tab:charm+_structure},\ref{tab:charm-_structure},\ref{tab:bottom+_structure},\ref{tab:bottom-_structure}, we can see that the different rms radii of the narrow resonances with  $ J^P=1^+ $ are of the same order and the differences between them are rather small. Therefore, the color mixing matrix elements are suppressed in the system.
	
	For the $ 1^{+-} $ $ cc\bar c\bar c $ system, the state $ T_{4c,1(-)}(6932) $ is a narrow resonance, whose dominant color-spin configuration is $ [(c\bar c)_{8_c}^{1}(c\bar c)_{8_c}^{0}]_{1_c}^{1} $. On the other hand, the dicharmonium thresholds $ \eta_c\psi $ have color-spin configuration $ [(c\bar c)_{1_c}^{1}(c\bar c)_{1_c}^{0}]_{1_c}^{1} $. The transition between $ T_{4c,1(-)}(6932) $ and the dicharmonium channels can only occur via color mixing, which is suppressed in the fully heavy system. For the $ 1^{++} $ $ cc\bar c\bar c $ system, there are two types of narrow resonances. The first type $ T_{4c,1(+)}(6896) $ and $ T_{4c,1(+)}(7200) $ have dominant color-spin configuration $ [(c\bar c)_{8_c}^{1}(c\bar c)_{8_c}^{1}]_{1_c}^{1} $ , while the second type $ T_{4c,1(+)}(6939) $ and $ T_{4c,1(+)}(7242) $ have dominant color-spin configuration $ [(cc)_{\bar3_c}^{1}(\bar c\bar c)_{3_c}^{1}]_{1_c}^1 $. On the other hand, the dicharmonium thresholds $ \Psi(2S)J/\psi $, $ \Psi(3S)J/\Psi $ have color-spin configuration $ [(c\bar c)_{1_c}^{1}(c\bar c)_{1_c}^{1}]_{1_c}^{1} $. The coupling between the first type of resonances and the docharmonium channels is suppressed by the color mixing matrix elements. The spin configurations of the second type of resonances are orthogonal to those of the dicharmonium channels, which can be seen from the decomposition,
     \begin{equation}
     \begin{aligned}
    [(c_1c_2)^1(\bar c_3\bar c_4)^1]^1=&\frac{1}{\sqrt{2}}[(c_1\bar c_3)^1(c_2\bar c_4)^0]^1+\\
    &\frac{1}{\sqrt{2}}[(c_1\bar c_3)^0(c_2\bar c_4)^1
]
^1.
    \end{aligned}
     \end{equation}
    As a result, the coupling between them can only occur via the color magnetic term. Therefore the two-body decay widths of these two types of resonant states are both suppressed. For the $ bb\bar b\bar b $ system, similar arguments can be used to account for the narrow widths of the states $ T_{4b,1(-)}(19762) $, $ T_{4b,1(+)}(19731) $ and $ T_{4b,1(+)}(19748) $.

	\section{Summary}\label{sec:summary}
	In summary, we calculate the mass spectrum of the S-wave fully heavy tetraquark systems with both normal $ (J^{PC}=0^{++},1^{+-},2^{++}) $ and exotic $ (J^{PC}=0^{+-},1^{++},2^{+-}) $ C-parities  using three different quark potential models (AL1, AP1, BGS). The exotic C-parity systems refer to the ones that have no corresponding S-wave ground  heavy quarkonia thresholds. We employ the Gaussian expansion method to solve the four-body Schrödinger equation, and the complex scaling method to distinguish resonant states from scattering states.
	
\begin{figure*}[tbp]
		\centering
		\includegraphics[width=1\linewidth]{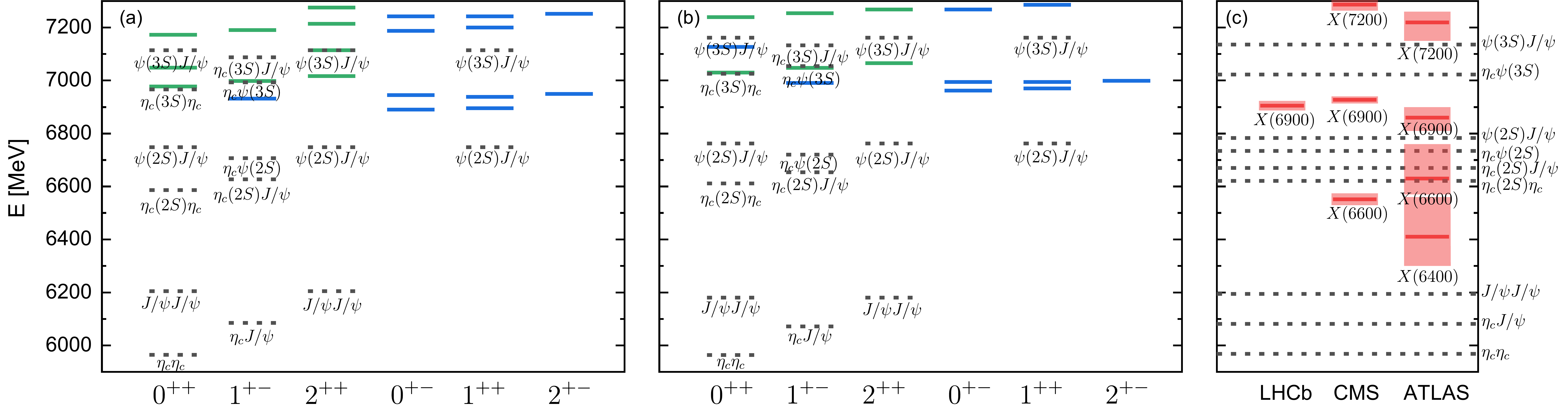}
		\caption{The mass spectrum of the S-wave $ cc\bar c\bar c $ states in the (a) AP1, (b) BGS potential. The experimental results reported by LHCb (model \uppercase\expandafter{\romannumeral1})~\cite{LHCb:2020bwg}, CMS (noninterference model)~\cite{CMS:2023owd} and ATLAS (model A and $ \alpha $)~\cite{ATLAS:2023bft} are shown in Fig.~(c).  In Fig.~(a) and (b), the green and blue lines represent the theoretical $ cc\bar c\bar c $ states with widths larger and smaller than $ 1\,\mathrm{MeV} $, respectively. In Fig.~(c), the red lines and the dashed areas represent the central masses and the uncertainties in the experiments, respectively. The dotted lines represent dicharmonium thresholds, whose experimental energies are taken from Ref.~\cite{ParticleDataGroup:2022pth}. }
		\label{fig:spectrum4c}
	\end{figure*}

\begin{figure}[tbp]
		\centering
		\includegraphics[width=0.85\linewidth]{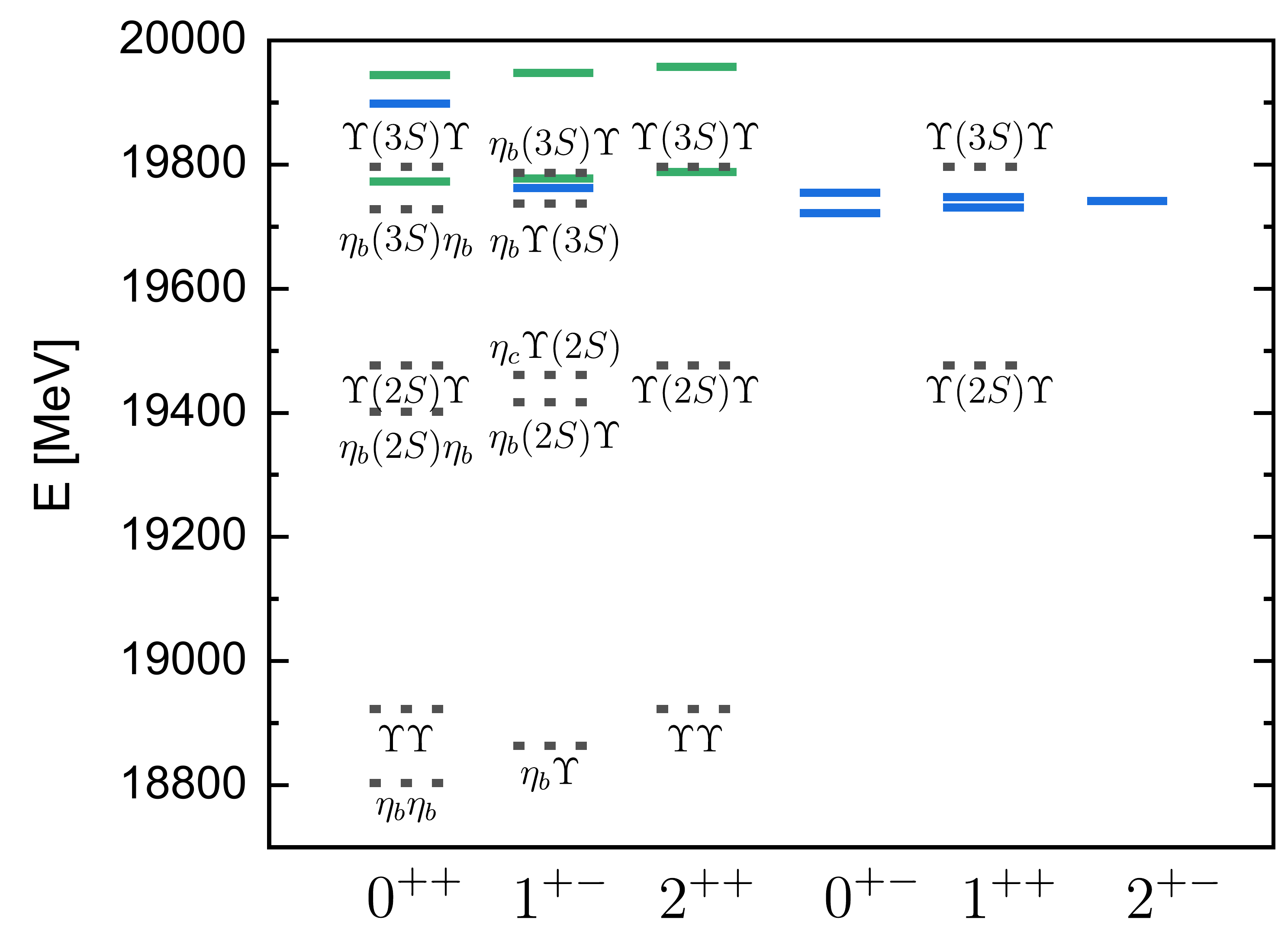}
		\caption{The mass spectrum of the S-wave $ bb\bar b\bar b $ states in the AP1 potential. The green and blue lines represent the states with widths larger and smaller than $ 1\,\mathrm{MeV} $, respectively. The dotted lines represent dicharmonium thresholds, whose experimental energies are taken from Ref.~\cite{ParticleDataGroup:2022pth}. }
		\label{fig:spectrum4b}
	\end{figure}

	Our calculations show that the mass spectra in different quark models are in qualitative agreement. We obtain a series of resonant states with $ J^{PC}=0^{++},1^{+-},2^{++
}$ and $ 1^{++} $. Moreover, we obtain several zero-width states in the $ 0^{+-} 
$ and $ 2^{
+-} $ systems, where S-wave diquarkonium threshold does not exist.  For the fully charmed system, we compare the theoretical results in the AP1 and BGS potentials with the experimental results in Fig.~\ref{fig:spectrum4c}.  We do not display the results in the AL1 potential since they are nearly the same as those in the AP1 potential.  We find good candidates for the experimentally observed $ X(6900) $ and $ X(7200) $ in both the $ 0^{++} $ and $ 2^{++} $ systems. However, signals for the $ X(6400) $ and $ X(6600) $ are not seen in our calculations. Several resonant and zero-width fully charmed tetraquark states await to be found in the experiment. For the fully bottomed system, we summarize the theoretical results in the AP1 potential in Fig.~\ref{fig:spectrum4b}. Resonant and zero-width fully bottomed tetraquark states are predicted in the mass region $(19.7,20.0)
\, \mathrm{GeV}$.
	
	By investigating the root mean square radii of the states, we find that most of the resonant and zero-width states have compact tetraquark configuration, except that the $ X(7200) $ candidates $ T_{4c,0(+)}(7173), T_{4c,2(+)}(7214) $ and the state $ T_{4c,1(-)}(7191) $ may have molecular configurations. We also study the decay modes of the resonant and zero-width states. The resonant states can decay strongly to S-wave diquarkonium thresholds while the zero-width states can only decay to P-wave quarkonia. Further study that considers the P-wave tetraquark systems is needed to better establish the properties of the $ 0^{+-} $ and $ 2^{+-} $ states.
 
	\section*{ACKNOWLEDGMENTS}
	
	We thank Zi-Yang Lin, Jun-Zhang Wang, Yao Ma, and Liang-Zhen Wen for the helpful discussions. This project was supported by the National
	Natural Science Foundation of China (11975033 and 12070131001). This
	project was also funded by the Deutsche Forschungsgemeinschaft (DFG,
	German Research Foundation, Project ID 196253076-TRR 110). The computational resources were supported by High-performance Computing Platform of Peking University.
	
	\appendix
	\section{COLOR BASIS AND COLOR FACTORS}\label{app:color}
	The color basis of the tetraquark systems can be written in the diquark-antidiquark form,
	\begin{equation}\label{appeq:color_diq}
		\begin{aligned}
			\chi_{\bar 3_c\otimes 3_c}=\left[\left(Q_1Q_2\right)_{\bar 3_c}\left(\bar Q_3\bar Q_4\right)_{3_c}\right]_{1_c},\\
			\chi
_{6_c\otimes \bar 6_c}=\left[\left(Q_1Q_2\right)_{6_c}\left(\bar Q_3\bar Q_4\right)_{\bar 6_c}\right]_{1_c},\\
		\end{aligned}
	\end{equation}
	or in two sets of dimeson form, 
	\begin{equation}\label{appeq:color_dimesona}
		\begin{aligned}
			\chi^a_{1_c\otimes 1_c}=\left[\left(Q_1\bar Q_3\right)_{1_c}\left(Q_2\bar Q_4\right)_{1_c}\right]_{1_c},\\
			\chi^a_{8_c\otimes 8_c}=\left[\left(Q_1\bar Q_3\right)_{8_c}\left(Q_2\bar Q_4\right)_{8_c}\right]_{1_c},\\
		\end{aligned}
	\end{equation}
	\begin{equation}\label{appeq:color_dimesonb}
		\begin{aligned}
			\chi^b_{1_c\otimes 1_c}=\left[\left(Q_1\bar Q_4\right)_{1_c}\left(Q_2\bar Q_3\right)_{1_c}\right]_{1_c},\\
			\chi^b_{8_c\otimes 8_c}=\left[\left(Q_1\bar Q_4\right)_{8_c}\left(Q_2\bar Q_3\right)_{8_c}\right]_{1_c}.\\
		\end{aligned}
	\end{equation}
	Each of these forms constitute a complete and orthogonal color basis for the tetraquark systems. The transformation between different sets of bases is given by
	\begin{equation}\label{appeq:color_tran}
		\begin{aligned}
			&\chi^a_{1_c\otimes1_c}=\frac{1}{\sqrt{3}}(\chi_{\bar 3_c\otimes 3_c}+\sqrt{2}\chi_{6_c\otimes \bar6_c}),\\
			&\chi^a_{8_c\otimes8_c}=\frac{1}{\sqrt{3}}(-\sqrt{2}\chi_{\bar 3_c\otimes 3_c}+\chi_{6_c\otimes \bar6_c}),\\
			&\chi^b_{1_c\otimes1_c}=\frac{1}{\sqrt{3}}(-\chi_{\bar 3_c\otimes 3_c}+\sqrt{2}\chi_{6_c\otimes \bar6_c}),\\
			&\chi^b_{8_c\otimes8_c}=\frac{1}{\sqrt{3}}(\sqrt{2}\chi_{\bar 3_c\otimes 3_c}+\chi_{6_c\otimes \bar6_c}).\\
		\end{aligned}
	\end{equation}
	It is equivalent to use Eqs.~\eqref{appeq:color_diq}, \eqref{appeq:color_dimesona} or \eqref{appeq:color_dimesonb} as the color basis in the calculations. The matrix elements for the color factor $\boldsymbol\lambda_i \cdot \boldsymbol\lambda_j$ in the diquark-antidiquark color basis~\eqref{appeq:color_diq} are listed in Table~\ref{tab:color_fac}. From the last column we can see that $ \sum_{i<j=1}^4\boldsymbol\lambda_i\cdot\boldsymbol\lambda_j $ is actually proportional to the identity operator.
	
	\begin{table*}[htbp]
		\centering
		\caption{Color factor matrix elements $\langle \chi_c |\boldsymbol{\lambda}_i \cdot \boldsymbol{\lambda}_j |\chi_c^\prime\rangle$ in the color basis~\eqref{appeq:color_diq}.}
		\label{tab:color_fac}
		\begin{tabular*}{\hsize}{@{}@{\extracolsep{\fill}}cccccccc@{}}
			\hline\hline
			$\langle \chi_c |\bm{\lambda}_i \cdot \bm{\lambda}_j |\chi_c^{\prime}\rangle$ & $\bm{\lambda}_1 \cdot \bm{\lambda}_2$ & $\bm{\lambda}_3 \cdot \bm{\lambda}_4$ & $\bm{\lambda}_1 \cdot \bm{\lambda}_3$ & $\bm{\lambda}_1 \cdot \bm{\lambda}_4$ & $\bm{\lambda}_2 \cdot \bm{\lambda}_3$ & $\bm{\lambda}_2 \cdot \bm{\lambda}_4$&$ \sum_{i<j=1}^4\boldsymbol\lambda_i\cdot\boldsymbol\lambda_j $  \\ 
			\hline
			$\langle \chi_{\bar3_c\otimes 3_c} |\bm{\lambda}_i \cdot \bm{\lambda}_j |\chi_{\bar3_c\otimes 3_c}\rangle$ & $-\frac{8}{3}$ & $-\frac{8}{3}$ & $-\frac{4}{3}$ & $-\frac{4}{3}$ & $-\frac{4}{3}$ & $-\frac{4}{3}$&-$\frac{32}{3}$  \\ 
			$\langle \chi_{6_c\otimes \bar6_c} |\bm{\lambda}_i \cdot \bm{\lambda}_j |\chi_{6_c\otimes \bar6_c}\rangle$ & $\frac{4}{3}$ & $\frac{4}{3}$ & $-\frac{10}{3}$ & $-\frac{10}{3}$ & $-\frac{10}{3}$ & $-\frac{10}{3}$& -$\frac{32}{3}$ \\ 
			
			$\langle \chi_{\bar3_c\otimes 3_c} |\bm{\lambda}_i \cdot \bm{\lambda}_j |\chi_{6_c\otimes \bar6_c}\rangle$ & $0$ & $0$ & $-2\sqrt{2}$ & $2\sqrt{2}$ & $2\sqrt{2}$ & $-2\sqrt{2}$&0  \\ \hline\hline
		\end{tabular*}
	\end{table*}
	
	\section{BASIS FUNCTIONS WITH DEFINITE C-PARITY}\label{app:basis}
	Under charge conjugation, the transformation behavior of the basis functions is shown in Eqs.~\eqref{eq:Ctransform1}-\eqref{eq:Ctransform3}.
	Using the linear superposition of the original basis functions, we can construct a new set of basis with definite C-parity and decompose the Hilbert space $ \mathcal{H} $ into positive and negative C-parity subspaces $ \mathcal{H}_\pm $. The basis functions of $ \mathcal{H}_\pm $ with different quantum numbers and satisfying the antisymmetrization of identical particles are explicitly listed in the following. 
	\begin{itemize}
		\item $ J^{PC}=0^{++} $
		\begin{align}
			&\chi^{0}_{\bar 3_c\otimes 3_c,1,1}[\Phi^{(\rm{c})}_{n_1,n_2}+\Phi^{(\rm{c})}_{n_2,n_1}] ,\\
			&\chi^{0}_{6_c\otimes\bar6_c,0,0}[\Phi^{(\rm{c})}_{n_1,n_2}+\Phi^{(\rm{c})}_{n_2,n_1}] ,\\
			&\chi^{0}_{\bar 3_c\otimes 3_c,1,1}[\Phi^{(\rm{a})}_{n_{1},n_{2}}+\Phi^{(\rm{b})}_{n_{1},n_{2}}+\Phi^{(\rm{a})}_{n_{2},n_{1}}+\Phi^{(\rm{b})}_{n_{2},n_{1}}] ,\\
			&\chi^{0}_{6_c\otimes\bar6_c,0,0}[\Phi^{(\rm{a})}_{n_{1},n_{2}}+\Phi^{(\rm{b})}_{n_{1},n_{2}}+\Phi^{(\rm{a})}_{n_{2},n_{1}}+\Phi^{(\rm{b})}_{n_{2},n_{1}}] ,\\
			&\chi^{0}_{\bar 3_c\otimes 3_c,0,0}[\Phi^{(\rm{a})}_{n_{1},n_{2}}-\Phi^{(\rm{b})}_{n_{1},n_{2}}+\Phi^{(\rm{a})}_{n_{2},n_{1}}-\Phi^{(\rm{b})}_{n_{2},n_{1}}] ,\\
			&\chi^{0}_{6_c\otimes\bar6_c,1,1}[\Phi^{(\rm{a})}_{n_{1},n_{2}}-\Phi^{(\rm{b})}_{n_{1},n_{2}}+\Phi^{(\rm{a})}_{n_{2},n_{1}}-\Phi^{(\rm{b})}_{n_{2},n_{1}}] ,
		\end{align}
		
		\item $ J^{PC}=0^{+-} $
		
		\begin{align}
			&\chi^{0}_{\bar 3_c\otimes 3_c,1,1}[\Phi^{(\rm{c})}_{n_1,n_2}-\Phi^{(\rm{c})}_{n_2,n_1}] ,\label{appeq1}\\
			&\chi^{0}_{6_c\otimes\bar6_c,0,0}[\Phi^{(\rm{c})}_{n_1,n_2}-\Phi^{(\rm{c})}_{n_2,n_1}] \label{appeq2},
		\end{align}
		
		\item $ J^{PC}=1^{++} $
		\begin{align}
			&\chi^{1}_{\bar 3_c\otimes 3_c,1,1}[\Phi^{(\rm{c})}_{n_1,n_2}-\Phi^{(\rm{c})}_{n_2,n_1}] ,\\
			&\chi^{1}_{\bar 3_c\otimes 3_c,1,0}[\Phi^{(\rm{a})}_{n_{1},n_{2}}-\Phi^{(\rm{b})}_{n_{1},n_{2}}-\Phi^{(\rm{a})}_{n_{2},n_{1}}+\Phi^{(\rm{b})}_{n_{2},n_{1}}]\notag\\
			+&\chi^{1}_{\bar 3_c\otimes 3_c,0,1}[\Phi^{(\rm{a})}_{n_{1},n_{2}}+\Phi^{(\rm{b})}_{n_{1},n_{2}}-\Phi^{(\rm{a})}_{n_{2},n_{1}}-\Phi^{(\rm{b})}_{n_{2},n_{1}}] ,\\
			&\chi^{1}_{6_c\otimes\bar6_c,0,1}[\Phi^{(\rm{a})}_{n_{1},n_{2}}-\Phi^{(\rm{b})}_{n_{1},n_{2}}-\Phi^{(\rm{a})}_{n_{2},n_{1}}+\Phi^{(\rm{b})}_{n_{2},n_{1}}]\notag\\
			+&\chi^{1}_{6_c\otimes\bar6_c,1,0}[\Phi^{(\rm{a})}_{n_{1},n_{2}}+\Phi^{(\rm{b})}_{n_{1},n_{2}}-\Phi^{(\rm{a})}_{n_{2},n_{1}}-\Phi^{(\rm{b})}_{n_{2},n_{1}}] ,
		\end{align}
		
		\item $ J^{PC}=1^{+-} $
		\begin{align}
			&\chi^{1}_{\bar 3_c\otimes 3_c,1,1}[\Phi^{(\rm{c})}_{n_1,n_2}+\Phi^{(\rm{c})}_{n_2,n_1}] ,\\
			&\chi^{1}_{\bar 3_c\otimes 3_c,1,1}[\Phi^{(\rm{a})}_{n_{1},n_{2}}+\Phi^{(\rm{b})}_{n_{1},n_{2}}+\Phi^{(\rm{a})}_{n_{2},n_{1}}+\Phi^{(\rm{b})}_{n_{2},n_{1}}],\\
			&\chi^{1}_{6_c\otimes\bar6_c,1,1}[\Phi^{(\rm{a})}_{n_{1},n_{2}}-\Phi^{(\rm{b})}_{n_{1},n_{2}}+\Phi^{(\rm{a})}_{n_{2},n_{1}}-\Phi^{(\rm{b})}_{n_{2},n_{1}}],\\
			&\chi^{1}_{\bar 3_c\otimes 3_c,1,0}[\Phi^{(\rm{a})}_{n_{1},n_{2}}-\Phi^{(\rm{b})}_{n_{1},n_{2}}-\Phi^{(\rm{a})}_{n_{2},n_{1}}+\Phi^{(\rm{b})}_{n_{2},n_{1}}]\notag\\
			-&\chi^{1}_{\bar 3_c\otimes 3_c,0,1}[\Phi^{(\rm{a})}_{n_{1},n_{2}}+\Phi^{(\rm{b})}_{n_{1},n_{2}}-\Phi^{(\rm{a})}_{n_{2},n_{1}}-\Phi^{(\rm{b})}_{n_{2},n_{1}}] ,\\
			&\chi^{1}_{6_c\otimes\bar6_c,0,1}[\Phi^{(\rm{a})}_{n_{1},n_{2}}-\Phi^{(\rm{b})}_{n_{1},n_{2}}-\Phi^{(\rm{a})}_{n_{2},n_{1}}+\Phi^{(\rm{b})}_{n_{2},n_{1}}]\notag\\
			-&\chi^{1}_{6_c\otimes\bar6_c,1,0}[\Phi^{(\rm{a})}_{n_{1},n_{2}}+\Phi^{(\rm{b})}_{n_{1},n_{2}}-\Phi^{(\rm{a})}_{n_{2},n_{1}}-\Phi^{(\rm{b})}_{n_{2},n_{1}}],
		\end{align}
		
		\item $ J^{PC}=2^{++} $
		\begin{align}
			&\chi^{2}_{\bar 3_c\otimes 3_c,1,1}[\Phi^{(\rm{c})}_{n_1,n_2}+\Phi^{(\rm{c})}_{n_2,n_1}] ,\\
			&\chi^{2}_{\bar 3_c\otimes 3_c,1,1}[\Phi^{(\rm{a})}_{n_{1},n_{2}}+\Phi^{(\rm{b})}_{n_{1},n_{2}}+\Phi^{(\rm{a})}_{n_{2},n_{1}}+\Phi^{(\rm{b})}_{n_{2},n_{1}}] ,\\
			&\chi^{2}_{6_c\otimes\bar6_c,1,1}[\Phi^{(\rm{a})}_{n_{1},n_{2}}-\Phi^{(\rm{b})}_{n_{1},n_{2}}+\Phi^{(\rm{a})}_{n_{2},n_{1}}-\Phi^{(\rm{b})}_{n_{2},n_{1}}] ,
		\end{align}
		
		\item $ J^{PC}=2^{+-} $
		\begin{align}
			&\chi^{2}_{\bar 3_c\otimes 3_c,1,1}[\Phi^{(\rm{c})}_{n_1,n_2}-\Phi^{(\rm{c})}_{n_2,n_1}] \label{appeq3},
		\end{align}
	\end{itemize}
	where $ \chi_{c_1\otimes c_2,s_1,s_2}^J $ is the color-spin wave function, and $ \Phi^{(\rm{jac})}_{n_1,n_2}\equiv \Phi^{(\rm{jac})}_{n_1,n_2,n_3}$ is the S-wave Gaussian spatial wave function. The detail expression for the wave functions can be found in Eqs.~\eqref{eq:colorspin_wf1} and \eqref{eq:spatial_wf}.
	
	We can see that the S-wave fully heavy tetraquark states with $ J^{PC}=0^{+-} $ and $ 2^{+-} $ contain only diquark-antidiquark configuration $ \Phi^{(\rm{c})}_{n_1,n_2,n_3} $ and exclude dimeson configuration $ \Phi^{(\rm{a})}_{n_1,n_2,n_3} $,  $  \Phi^{(\rm{b})}_{n_1,n_2,n_3} $, which arises from the fact that S-wave dimeson systems cannot have such quantum numbers. It should be noted that the minus sign in Eqs.~\eqref{appeq1}, \eqref{appeq2} and \eqref{appeq3} demands that the radial excitation of the diquark and the antidiquark must be different, therefore the $ 0^{+-} $ or $ 2^{+-} $ tetraquark state is not a particle-antiparticle pair and one cannot simply calculate the C-parity as  $ C=(-1)^{L+S}=+1 $. 

 \section{TWO DEFINITIONS OF ROOT MEAN SQUARE RADIUS }\label{app:rms}
\clabel[q11]{In our calculations, we only use the non-antisymmetric component of the wave function to calculate the rms radii. It seems more reasonable to calculate the rms radii of the compact tetraquark states using the complete wave function. However, our primary interest lies in the general clustering behavior of the tetraquark states rather than in specific numerical results of the rms radii, since the latter are not experimentally observables at present. The rms radii calculated using the non-antisymmetric component of the wave function are already capable of distinguishing between compact tetraquark states and loose molecular states.  To illustrate this, we compare the results of rms radii calculated using the complete wave function $|\Psi^J(\theta)\rangle $ and the non-antisymmetric term $|\Psi^J_{\rm nA}(\theta)\rangle $ in Table~\ref{tab:rms}. One compact and one molecular $cc\bar c\bar c$ resonant states with $J^{PC}=0^{++}$ are chosen as examples. For the state $T_{4c,0(+)}(6978)$, both the results from $|\Psi^J(\theta)\rangle $ and $|\Psi^J_{\rm nA}(\theta)\rangle $ indicate that all rms radii are of the same order and support the compact tetraquark configuration. For the state $T_{4c,0(+)}(7173)$, the results from $|\Psi^J_{\rm nA}(\theta)\rangle$ can clearly demonstrate the clustering behaviour of a molecular state: $(c_1\bar c_3)$ and $(c_2\bar c_4)$ form two subclusters, which are widely separated. On the other hand, the results from  $|\Psi^J(\theta)\rangle$ are much less clear due to the antisymmetrization. It should also be noted that all $r_{c_i\bar c_j}^{\rm rms}$ are the same in the results from  $|\Psi^J(\theta)\rangle$ due to the antisymmetrization of the wave function. }
\begin{table*}[htp]
\centering
\caption{\change{The rms radii (in fm) of the $ cc\bar c\bar c $ resonant states calculated using the complete wave functions $|\Psi^J(\theta)\rangle $ in Eq.~\eqref{eq:wf_decompose} and the non-antisymmetric components of the wave functions $|\Psi^J_{\rm nA}(\theta)\rangle $ in Eq.~\eqref{eq:nAwf}, respectively. The last column shows the spatial configurations of the states, where C. and M. represent the compact tetraquark and molecular configurations, respectively. }}
\label{tab:rms}
\begin{tabular}{ccccccc}
    \hline\hline
    $  M-i\Gamma/2 $ &Wave function& $ r_{c_1\bar{c}_3}^{\mathrm{rms}} $&$ r_{c_2\bar{c}_4}^{\mathrm{rms}} $&$ r_{c_1\bar{c}_4}^{\mathrm{rms}} $=$ r_{c_2\bar{c}_3}^{\mathrm{rms}} $&$ r_{c_1c_2}^{\mathrm{rms}} $=$ r_{\bar{c}_3\bar{c}_4}^{\mathrm{rms}} $& Configurations\\
    \hline
    
    $ 6978-36i $&  $|\Psi^J_{\rm nA}(\theta)\rangle $&$0.81$&$0.81$&$0.86$&$0.66$&C.\\
    &$|\Psi^J(\theta)\rangle $&$0.83$&$0.83$&$0.83$&$0.68$&\\
    $ 7173-20i $&$|\Psi^J_{\rm nA}(\theta)\rangle $&$0.89$&$0.89$&$2.31$&$2.28$&M.\\
    &$|\Psi^J(\theta)\rangle $&$1.83$&$1.83$&$1.83$&$2.43$&\\
    \hline\hline
			
	\end{tabular}
\end{table*}

\change{In conclusion, calculating the rms radii using only the non-antisymmetric term can reflect the internal spatial structure of tetraquark states more transparently. }
	
	\bibliography{QQQQref}
	
	\onecolumngrid
	\clearpage
	\twocolumngrid

\end{document}